\documentclass[11pt]{article}
\usepackage{cite}
\usepackage{mdframed}
\usepackage{epsfig,subfigure}
\usepackage{amssymb}
\usepackage[fleqn]{amsmath}
\usepackage{color}
\usepackage{arydshln}
\def\QED{\mbox{\rule[0pt]{1.5ex}{1.5ex}}}

\usepackage{graphicx}
\usepackage{color}
\usepackage[dvipsnames]{xcolor}

\definecolor{armygreen}{rgb}{0.29, 0.33, 0.13}

\usepackage{amssymb}
\usepackage{amsmath,amsfonts,amssymb}
\usepackage{verbatim}
\usepackage{stfloats}
\usepackage[bookmarks=false]{}
\usepackage{algorithm}
\usepackage{algcompatible}

\newtheorem{theorem}{Theorem}

\newtheorem{lemma}{Lemma}

\newtheorem{remark}{Remark}

\newcommand\blfootnote[1]{%
  \begingroup
  \renewcommand\thefootnote{}\footnote{#1}%
  \addtocounter{footnote}{-1}%
  \endgroup
}

\begin{document}
\date{}

\title{
Information Theoretic Secure Aggregation with User Dropouts
}
\author{\normalsize Yizhou Zhao and Hua Sun \\
}

\maketitle

\blfootnote{
Yizhou Zhao (email: yizhouzhao@my.unt.edu) and Hua Sun (email: hua.sun@unt.edu) are with the Department of Electrical Engineering at the University of North Texas. }

\maketitle

\begin{abstract}
In the robust secure aggregation problem, a server wishes to learn and only learn the sum of the inputs of a number of users while some users may drop out (i.e., may not respond). The identity of the dropped users is not known a priori and the server needs to securely recover the sum of the remaining surviving users. We consider the following minimal two-round model of secure aggregation. Over the first round, any set of no fewer than $U$ users out of $K$ users respond to the server and the server wants to learn the sum of the inputs of all responding users. The remaining users are viewed as dropped. Over the second round, any set of no fewer than $U$ users of the surviving users respond (i.e., dropouts are still possible over the second round) and from the information obtained from the surviving users over the two rounds, the server can decode the desired sum. The security constraint is that even if the server colludes with any $T$ users and the messages from the dropped users are received by the server (e.g., delayed packets), the server is not able to infer any additional information beyond the sum in the information theoretic sense. For this information theoretic secure aggregation problem, we characterize the optimal communication cost. When $U \leq T$, secure aggregation is not feasible, and when $U > T$, to securely compute one symbol of the sum, the minimum number of symbols sent from each user to the server is $1$ over the first round, and $1/(U-T)$ over the second round.
\end{abstract}

\newpage

\allowdisplaybreaks
\section{Introduction}
The rapidly increasing volume of data available at massive distributed nodes enables powerful large-scale learning applications. For example, in federated learning \cite{federated, federated_survey, federated_survey1}, a large number of mobile users wish to collaboratively train a shared global model, coordinated by a central server. While the distributed users are willing to cooperate with the server to learn the shared model, they do not fully trust the server and do not want to reveal any information beyond what is necessary to train the desired model. 
Specifically, 
when the local models of the distributed users are aggregated (in the form of summation usually) at the server to produce the global model, each user does not want to reveal any additional information about its local data. Therefore, regarding security, the central technical problem is secure sum computation or secure aggregation \cite{user_held, aggregation}, i.e., how to compute, with as little communication as possible, the sum of the inputs of a number of users without exposing any information beyond the sum. A particular challenge in secure aggregation brought by federated learning is the phenomenon of user dropouts, i.e., some users whose identities are not known beforehand may drop from the learning procedure (due to unreliable communication connections or limited battery life) and the server needs to be able to robustly recover the sum of the inputs of the remaining surviving users while learning nothing else at the same time. The robustness to dropped users is a key requirement that calls for novel models and analysis. 
The main objective of this work is to understand the fundamental communication limits of information theoretic secure aggregation with user dropouts. 

\subsection*{Secure Aggregation with User Dropouts}

The secure aggregation problem is comprised of one server and $K$ users. User $k, k \in \{1,2,\cdots, K\}$ holds an input $W_k$, which is a vector of $L$ elements from a field. In federated learning, the input $W_k$ may represent the local model, model update, gradient, loss, or parameters of User $k$, from one iteration of the iterative training optimization process and is typically high-dimensional, i.e., $L$ is large, which matches well with the Shannon theoretic formulation where $L$ is allowed to approach infinity. 
In this work, we focus on such inputs from one iteration as the secure aggregation problem remains the same for all iterations. 
 A randomness variable $Z_k$, independent of all inputs, is generated offline (before the values of $W_1, \cdots, W_K$ are known) and is available to User $k$ to assist with the secure aggregation task.

The server wishes to compute the element-wise sum of the vector inputs of all users. To do so, each user sends a message $X_k$, as a function of $W_k$ and $Z_k$, to the server. However, due to user dropouts, the server may not receive all messages; if only the messages from the set of users $\mathcal{U}_1$ arrive at the server and other messages are dropped, then the server wants to securely compute $\sum_{k \in \mathcal{U}_1} W_k$, i.e., the sum of the inputs of all responding users, from $(X_k : k \in \mathcal{U}_1)$. For example, suppose $K = 4$ and $\mathcal{U}_1 = \{1,2,3\}$. Then the server sees only $X_1, X_2, X_3$ and wants to recover $W_1 + W_2 + W_3$ while learning no other information, e.g., the server cannot infer $W_1 + W_2$. We now observe an inherent deficiency of such a model, caused by the uncertainty of the identity of the dropped users. As it is not known a priori which users will drop, the sent messages $X_k$ cannot depend on the set of dropped users and must enable secure computation for all possible responding users. For example, if $\mathcal{U}_1 = \{1, 2\}$, then the server must be able to decode $W_1 + W_2$ from $X_1, X_2$, which contradicts the security constraint for the case where $\mathcal{U}_1 = \{1,2,3\}$, i.e., from $X_1, X_2, X_3$, the server can learn only $W_1 + W_2 + W_3$. 
Therefore, for the above communication model as the identity of the responding users is unknown beforehand, it is not feasible to learn only the sum of their inputs and nothing else.

\begin{figure}[h]
\begin{center}
\includegraphics[width= 6 in]{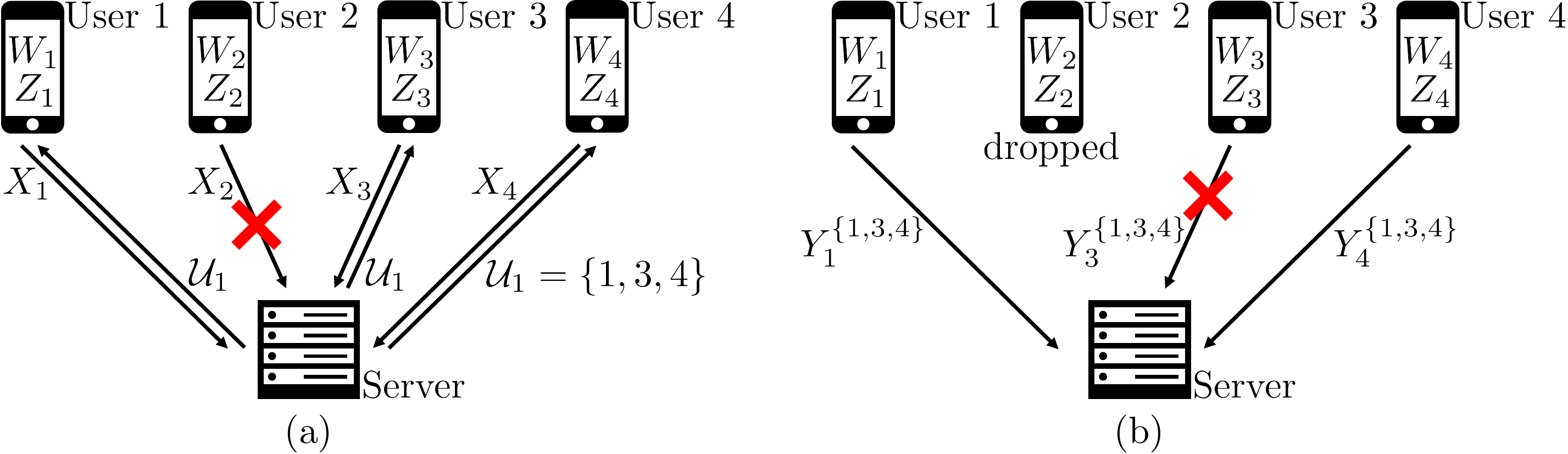}
\caption{\small A robust secure aggregation problem instance with $K=4$ users. (a). Over the first round, User~2 is dropped; (b). Over the second round, User~3 is dropped. The server securely computes $W_1 + W_3 + W_4$.}
\label{fig:model}
\end{center}
\end{figure}

The remedy is to include additional rounds of communication, and this solution has been taken in prior works on secure aggregation \cite{user_held, aggregation, aggregation_log, aggregation_turbo, aggregation_fast}. In this work, we consider the simplest model of two rounds. We refer to the round that is discussed above and parameterized by $X_1, \cdots, X_K$, as the first round. At the end of the first round, the server informs all responding users about the surviving user set $\mathcal{U}_1$ and the remaining users 
are viewed as dropped thus no further communication with them is requested. One additional round of messages are requested from the surviving users in $\mathcal{U}_1$. This round is referred to as the second round and the message from User $k \in \mathcal{U}_1$ is denoted as $Y_k^{\mathcal{U}_1}$, where the superscript highlights that the identity of the surviving users over the first round is known when the user decides the second round message (also as a function of $W_k$ and $Z_k$). User dropouts are still possible over the second round and we denote the set of responding users over the second round by $\mathcal{U}_2$, which is a subset of $\mathcal{U}_1$. We assume that $|\mathcal{U}_2|$, the cardinality of $\mathcal{U}_2$, is at least $U$, a pre-determined threshold parameter. That is, the server will wait for at least $U$ users, e.g., by setting up a proper time deadline. As $\mathcal{U}_2 \subset \mathcal{U}_1$, we have $|\mathcal{U}_1| \geq U$.
The setup of this $U$ parameter is interpreted as 
the worst case estimate of the number of surviving users, and is also to make the secure aggregation problem more interesting. See Figure~\ref{fig:model} for an example where $K = 4, U = 2$, and $\mathcal{U}_1 = \{1,3,4\}$, $\mathcal{U}_2 = \{1,4\}$. 

\vspace{0.1in}
\vspace{-0.1in}

After describing the communication model, we now proceed to state the two constraints of secure aggregation - correctness and security. 
\begin{itemize}
\item Correctness constraint: From only the messages received from the surviving users over the two rounds, the server can decode $\sum_{k \in \mathcal{U}_1} W_k$ with no error. For example, in Figure~\ref{fig:model}, it is required that $W_1+W_3+W_4$ can be recovered from $X_1, X_3, X_4, Y_1^{\{1,3,4\}}, Y_4^{\{1,3,4\}}$.
\item {\it Information theoretic} security constraint: From {\em all} the messages sent from the users over the two rounds (including those from dropped users as their packets may be merely delayed) and even if the server colludes with any set of at most $T$ users, the server cannot infer any additional information in the information theoretic sense about all inputs $W_1, W_2, \cdots, W_K$ beyond what is already known from the colluding user(s) and the desired sum. For example, suppose $T=1$ and the colluding user is User $4$ in Figure~\ref{fig:model}, then it is required that from all the messages $X_1, X_2, X_3, X_4, Y_1^{\{1,3,4\}}, Y_3^{\{1,3,4\}}, Y_4^{\{1,3,4\}}$ and colluding user's information $W_4, Z_4$, no information about $W_1, W_2, W_3, W_4$ is revealed, except $W_4$ and $W_1 + W_3 + W_4$. Specifically, while $W_1 + W_3$ can be obtained, nothing more about $W_1$ or $W_3$ can be learned.
\end{itemize}
Importantly, we emphasize that a feasible secure aggregation protocol must satisfy the correctness and security constraints for any first round responding user set $\mathcal{U}_1$ where $|\mathcal{U}_1| \geq U$, any second round responding user set $\mathcal{U}_2$ where $\mathcal{U}_2 \subset \mathcal{U}_1$ and $|\mathcal{U}_2| \geq U$, and any colluding user set $\mathcal{T}$ where $|\mathcal{T}| \leq T$. A secure aggregation protocol specifies a design of the messages $X_k, Y_k^{\mathcal{U}_1}$ and we are interested in characterizing the optimal communication efficiency, i.e., minimizing the number of symbols contained in the messages $X_k$ and $Y_k^{\mathcal{U}_1}$.

As a recap, our information theoretic secure aggregation formulation contains $3$ parameters, $K$ (the number of users), $U$ (a threshold parameter on the minimum number of responding users), and $T$ (a threshold parameter on the maximum number of colluding users). 
We assume that $1 \leq U \leq K-1$, so there may exist dropped users; otherwise $U = K$ and the problem becomes degraded as all users must respond. We also assume that $1\leq T \leq K-2$; 
otherwise $T = K-1$ or $K$, then when the colluding user set contains at least $K-1$ users, there is nothing to hide, as from the desired sum and $K-1$ inputs from the colluding users, the server can decode all $K$ inputs. For this model, our main goal is to answer the following question - {\em to compute one symbol of the desired sum function securely, what is the minimum number of symbols that must be sent from the users over the first round and over the second round, as a function of $K, U, T$}?

\subsection*{Summary of Results}
We obtain a complete answer to the above question, i.e., the exact characterization of the optimal communication efficiency of information theoretic secure aggregation. Specially, we show that 
\begin{itemize}
\item when $U \leq T$, 
secure aggregation is not infeasible in the information theoretic sense; 
\item when $U > T$, the minimum number of symbols that each user needs to send is $1$ symbol over the first round, and $1/(U-T)$ symbols over the second round, per symbol of desired sum.
\end{itemize}

The proofs of the above result are fairly standard. The protocol design uses and adapts elements that are frequently encountered in secure (sum) computation literature \cite{BGW, CCD, Chor_Kushilevitz, Kushilevitz_Rosen, Zhou_Sun_Fu} (see Section \ref{sec:ach}). The entropy based proof of impossibility claims uses Shannon's information theoretic security framework \cite{Shannon1949}, which will be adapted to robust secure aggregation (see Section \ref{sec:con}) and is conceptually similar to that in (symmetric) private information retrieval context \cite{Sun_Jafar_SPIR, Jia_Sun_Jafar, guo2020information, cheng2020capacity, wang2018e, wang2019symmetric}. While the optimal communication efficiency is established, a number of relevant problems remain widely open, e.g., the minimum randomness consumption (see Section \ref{sec:dis}).

Let us conclude the introduction section by summarizing the major differences between our work and existing works on secure aggregation for federated learning, which has attracted tremendous recent attention \cite{user_held, aggregation, aggregation_log, aggregation_turbo, aggregation_fast, bonawitz2019federated, Choi_Sohn_Han_Moon, pillutla2019robust, xu2019hybridalpha, beguier2020safer, so2020byzantine, elkordy2020secure, guo2020secure, alexandru2020private, lia2020privacy, truong2020privacy}. First of all, to the best of our knowledge, our work is the only one that considers information theoretic security, i.e., unconditional security based on statistical independence; while all prior works focus on cryptographic security, i.e., conditional security against computationally bounded adversaries. Second, we first define the system parameters (e.g., allowed user dropouts and collusions), and then study the fundamental limits (i.e., the best possible protocols) given the specified parameters; while most existing works first propose a specific protocol and then analyze its performance (e.g., allowed user dropouts and collusions). Last but not least, we assume that the randomness variables of certain joint distribution are distributed to the users by a trusted third-party before the communication protocol starts (i.e., offline); while most prior works jointly consider randomness generation/distribution and message transmission (i.e., online). We view randomness generation as a separate problem to be studied in a future work, e.g., how to efficiently generate and distribute the required correlated randomness variables.



\section{Problem Statement}\label{sec:model}
The secure aggregation problem involves a server and $K$ users, where $K \geq 2$ and User $k \in \{1,2,\cdots, K\} \triangleq [K]$ holds an input vector $W_k$ and a randomness variable $Z_k$. The input vectors $\left(W_k\right)_{k\in[K]}$ are independent. Each $W_k$ is an $L \times 1$ column vector and the $L$ elements are i.i.d. uniform symbols from the finite field\footnote{The uniformity and independence of the input vectors are required for the converse proof, but are not necessary for the achievability proof (see Remark \ref{remark:noind}). {\color{black} Non-identically distributed inputs and the modulo ring of integers (versus finite fields) will be discussed in Section \ref{sec:dis}. 
}} 
$\mathbb{F}_q$. $\left(W_k\right)_{k\in[K]}$ is independent of $\left(Z_k\right)_{k\in[K]}$. 
\begin{eqnarray}
&& H\left(\left(W_k\right)_{k\in[K]}, \left(Z_k\right)_{k\in[K]}\right) = \sum_{k\in[K]} H(W_k) + H\left(\left(Z_k\right)_{k\in[K]} \right), \label{h1}\\
&& H(W_k) = L ~(\mbox{in $q$-ary units}), \forall k \in [K]. \label{h2}
\end{eqnarray}
The communication protocol between the server and the users has two rounds. Over the first round, User $k$ sends a message $X_k, k \in [K]$ to the server. The message $X_k$ is a function of $W_k, Z_k$ and consists of $L_X$ symbols from $\mathbb{F}_q$.
\begin{eqnarray}
H(X_k | W_k, Z_k) = 0, \forall k \in [K]. \label{1x}
\end{eqnarray}
Some users may drop and the set of surviving users after the first round is denoted as $\mathcal{U}_1$, which can be any set of at least $U$ users, and $1 \leq U \leq K-1$. The server receives the messages $\left(X_k\right)_{k\in\mathcal{U}_1}$ and wishes to securely compute $\sum_{k\in\mathcal{U}_1} W_k$, where the vector summation is defined as the element-wise addition over $\mathbb{F}_q$. To do so, the server informs all surviving users about $\mathcal{U}_1$ and requests a second round of messages from them. The second round message sent from User $k \in \mathcal{U}_1$ is denoted as $Y_{k}^{\mathcal{U}_1}$, which is a function of $W_k, Z_k$ and consists of $L_Y$ symbols\footnote{In general, the message length $L_Y$ may depend on the user index $k$ and the first round surviving user set $\mathcal{U}_1$, while we assume that $L_Y$ is a constant in this work. Equivalently, we consider the maximum (worst case) message length $L_Y$ over all $k$ and $\mathcal{U}_1$. Similarly, $L_X$ is also a constant and represents the maximum first round message length.} from $\mathbb{F}_q$.
\begin{eqnarray}
H(Y_k^{\mathcal{U}_1} | W_k, Z_k) = 0, \forall k \in \mathcal{U}_1, \forall \mathcal{U}_1 \subset [K], |\mathcal{U}_1| \geq U. \label{2y}
\end{eqnarray}
Some users may drop and the set of surviving users after the second round is denoted as $\mathcal{U}_2$, where $\mathcal{U}_2 \subset \mathcal{U}_1$ and $|\mathcal{U}_2| \geq U$. Then the server receives the messages $\left(Y_k^{\mathcal{U}_1}\right)_{k\in\mathcal{U}_2}$ over the second round.

From the messages received from surviving users, the server must be able to decode the desired sum $\sum_{k\in\mathcal{U}_1} W_k$ with no error\footnote{Note that the results of this work also hold under vanishing error and leakage framework, i.e., when 0 is replaced by $o(L)$ in (\ref{corr}) and (\ref{sec}).}, i.e., the following correctness constraint must be satisfied for any $\mathcal{U}_1, \mathcal{U}_2$, where $\mathcal{U}_2 \subset \mathcal{U}_1 \subset{[K]}, |\mathcal{U}_2| \geq U$.
\begin{eqnarray}
\mbox{[Correctness]}~~~H\left(\sum_{k\in\mathcal{U}_1} W_k \Bigg| \left(X_k\right)_{k\in\mathcal{U}_1}, \left(Y_k^{\mathcal{U}_1}\right)_{k\in\mathcal{U}_2} \right) = 0. \label{corr}
\end{eqnarray}
We impose that security must be guaranteed even if the messages sent from all surviving {\em and dropped} users are received by the server and the server may collude with any set of at most $T$ users, where $1 \leq T \leq K-2$.
Specifically, security refers to the constraint that the server cannot infer any additional information about $\left(W_k\right)_{k\in[K]}$ beyond that contained in $\sum_{k\in\mathcal{U}_1} W_k$ and known from the colluding users. That is, the following security constraint must be satisfied for any $\mathcal{U}_1, \mathcal{T}$, where $\mathcal{U}_1, \mathcal{T} \subset{[K]}, |\mathcal{U}_1| \geq U, |\mathcal{T}|\leq T$.
\begin{eqnarray}
\mbox{[Security]}~~~I\left(\left(W_k\right)_{k\in[K]}; \left(X_k\right)_{k\in[K]}, \left(Y_k^{\mathcal{U}_1}\right)_{k\in\mathcal{U}_1} \Bigg| \sum_{k\in\mathcal{U}_1} W_k, \left( W_k, Z_k \right)_{k\in\mathcal{T}} \right) = 0.
\label{sec}
\end{eqnarray}
The communication {\em rate} characterizes how many symbols each message contains per input symbol, and is defined as follows.
\begin{eqnarray}
R_1 \triangleq \frac{L_X}{L}, ~R_2  \triangleq \frac{L_Y}{L} \label{rate}
\end{eqnarray}
where $R_1$ is the first round message rate and $R_2$ is the second round message rate. 

A rate tuple $(R_1, R_2)$ is said to be achievable if there exists a secure aggregation scheme (i.e., a design of the correlated randomness variables $\left(Z_k\right)_{k\in[K]}$ and the messages $\left(X_k\right)_{k\in[K]}, \left(Y_k^{\mathcal{U}_1}\right)_{k\in\mathcal{U}_1}$), for which the correctness and security constraints (\ref{corr}), (\ref{sec}) are satisfied, and the first round and second round message rates are smaller than or equal to $R_1$ and $R_2$, respectively. 
The closure of the set of all achievable rate tuples is called the optimal rate region, denoted as $\mathcal{R}^*$. 

\section{Main Result: Optimal Rate Region of Secure Aggregation}\label{sec:result}
Theorem \ref{thm:main} states the main result.

\begin{theorem}\label{thm:main}
For the information theoretic secure aggregation problem with $K$ users, at least $U$ responding users, and at most $T$ colluding users, where $1\leq U \leq K-1, 1\leq T \leq K-2$, the optimal rate region is
\begin{eqnarray}
\mathcal{R}^* = \left\{
\begin{array}{cl}
\emptyset & ~\mbox{when}~U \leq T, \\
 \left\{ \left(R_1, R_2\right) : R_1 \geq 1, R_2 \geq \frac{1}{U-T} \right\}& ~\mbox{when}~ U > T.
\end{array}
\right.
\end{eqnarray}
\end{theorem}

From Theorem \ref{thm:main} and its proof (see Section \ref{sec:ach} for achievability and Section \ref{sec:con} for converse), we have the following observations.
\begin{itemize}
\item When $U\leq T$, i.e., the minimum number of responding users is no greater than the maximum number of colluding users, the information theoretic secure aggregation problem is not feasible, i.e., it is not possible to simultaneously satisfy the correctness constraint (\ref{corr}) and the security constraint (\ref{sec}).
\item When $U > T$, the optimal communication-wise strategy is such that each user sends $1$ symbol over the first round, and $1/(U-T)$ symbols over the second round, for each input symbol (i.e., to compute one symbol of the desired sum). Note that the optimal rate does not depend on the number of users $K$, and it depends on $U, T$ only through their difference $U-T$. In particular, when the difference between the two threshold parameters $U-T$ is larger, fewer symbols need to be sent. While the optimal communication cost (per user) may not depend on $K$, the minimum randomness consumption (i.e., the entropy of each $Z_k$ and the joint entropy of $\left(Z_k\right)_{k\in[K]}$) depends on $K$ (see Section \ref{sec:dis}).
\item While the input length $L$ is allowed to approach infinity in the rate definition (\ref{rate}), the achievable scheme (presented in Section \ref{sec:ach}) only requires $L = U-T$ (or integer multiples of $U-T$) when the field size $q$ satisfies $q \geq K+U$, and for any field size, {\color{black} it suffices to have $L = B(U-T)$, where $B$ is any integer such that $q^B \geq K+U$}.
\end{itemize}

\section{Proof of Theorem \ref{thm:main}: Achievability}\label{sec:ach}
Before presenting the general achievability proof, we first consider two examples to illustrate the idea, which is fairly straightforward and relies on generic vector linear codes.

\subsection{Example 1: $K = 3, U = 2, T = 0$}
Consider $K=3$ users, where at least $U=2$ users will respond, and no user will collude with the server ($T=0$). 
Suppose the input length is $L = U-T = 2$, i.e., $W_k = (W_k(1); W_k(2)) \in \mathbb{F}_q^{2\times 1}$.

We first specify the randomness variables. Consider $3$ i.i.d. uniform $2 \times 1$ vectors over $\mathbb{F}_q$, denoted as $S_k = (S_k(1); S_k(2)), k \in \{1,2,3\}$ and yield generic linear combinations of the sum of all subsets of $\{S_1, S_2, S_3\}$ with cardinality no fewer than $U = 2$.
\begin{eqnarray}
~ \left[ \begin{array}{c}
Z_1^{\{1,2\}}\\
Z_2^{\{1,2\}}
\end{array}
\right] \triangleq \mbox{MDS}_{2\times 2} \left[ \begin{array}{c}
S_1(1) + S_2(1)\\
S_1(2) + S_2(2)
\end{array}
\right], &&
\left[ \begin{array}{c}
Z_1^{\{1,3\}}\\
Z_3^{\{1,3\}}
\end{array}
\right] \triangleq \mbox{MDS}_{2\times 2} \left[ \begin{array}{c}
S_1(1) + S_3(1)\\
S_1(2) + S_3(2)
\end{array}
\right], \notag \\
~ \left[ \begin{array}{c}
Z_2^{\{2,3\}}\\
Z_3^{\{2,3\}}
\end{array}
\right] \triangleq \mbox{MDS}_{2\times 2} \left[ \begin{array}{c}
S_2(1) + S_3(1)\\
S_2(2) + S_3(2)
\end{array}
\right], &&
\left[ \begin{array}{c}
Z_1^{\{1,2,3\}}\\
Z_2^{\{1,2,3\}}\\
Z_3^{\{1,2,3\}}
\end{array}
\right] \triangleq \mbox{MDS}_{3 \times 2} \left[ \begin{array}{c}
S_1(1) + S_2(1) + S_3(1)\\
S_1(2) + S_2(2) + S_3(2)
\end{array}
\right] \notag\\
&& \label{eq:mds}
\end{eqnarray}
where $\mbox{MDS}_{a\times b}$ denotes any MDS matrix of dimension $a \times b$. $\mbox{MDS}_{2\times 2}$ can be any full rank matrix, and it is presented using MDS matrices to facilitate generalizations to larger parameters. Note that the MDS matrices appeared in (\ref{eq:mds}) exist over any finite field. When $T > 0$, we require slightly stronger properties on theses matrices and will use Cauchy matrices (see the next section). Then we set
\begin{eqnarray}
Z_1 &=& \left(S_1, Z_1^{\{1,2\}}, Z_1^{\{1,3\}}, Z_1^{\{1,2,3\}} \right), \notag\\
Z_2 &=& \left(S_2, Z_2^{\{1,2\}}, Z_2^{\{2,3\}}, Z_2^{\{1,2,3\}} \right), \notag\\
Z_3 &=& \left(S_3, Z_3^{\{1,3\}}, Z_3^{\{2,3\}}, Z_3^{\{1,2,3\}} \right). \label{eq:zi}
\end{eqnarray}
We have completed the design of the correlated randomness variables. 

Next, we describe the design of the messages over two rounds. For the first round, we set
\begin{eqnarray}
X_1 = W_1 + S_1, ~X_2 = W_2 + S_2, ~X_3 = W_3 + S_3
\end{eqnarray}
where `$+$' denotes element-wise addition over $\mathbb{F}_q$. For the second round, we set
\begin{eqnarray}
\mathcal{U}_1 = \{1,2\}: && Y_1^{\{1,2\}} = Z_1^{\{1,2\}}, Y_2^{\{1,2\}} = Z_2^{\{1,2\}}, \notag\\
\mathcal{U}_1 = \{1,3\}: && Y_1^{\{1,3\}} = Z_1^{\{1,3\}}, Y_3^{\{1,3\}} = Z_3^{\{1,3\}}, \notag\\
\mathcal{U}_1 = \{2,3\}: && Y_2^{\{2,3\}} = Z_2^{\{2,3\}}, Y_3^{\{2,3\}} = Z_3^{\{2,3\}}, \notag\\
\mathcal{U}_1 = \{1,2,3\}: && Y_1^{\{1,2,3\}} = Z_1^{\{1,2,3\}}, Y_2^{\{1,2,3\}} = Z_2^{\{1,2,3\}}, Y_3^{\{1,2,3\}} = Z_3^{\{1,2,3\}}.
\end{eqnarray}

Finally, we prove that the scheme is correct and secure, and the rate tuple achieves the extreme point of the optimal rate region.

{\em Correctness:} 
When $\mathcal{U}_1 = \{1,2\}$, i.e., User 3 drops over the first round, we have $\mathcal{U}_2 = \mathcal{U}_1$ as $U=2$, i.e., at least $2$ users survive in the end, then no user drops over the second round. From the second round messages $Y_1^{\{1,2\}}, Y_2^{\{1,2\}}$, the server can recover $S_1 + S_2$ with no error, as the precoding matrices chosen in (\ref{eq:mds}) are MDS. Combining with the sum of the received two first round messages, $X_1 + X_2 = W_1 + W_2 + S_1 + S_2$, the desired sum $W_1 + W_2$ can be decoded with no error.
The correctness proof for other cases where $|\mathcal{U}_1| = 2$ follows similarly.

When $\mathcal{U}_1 = \{1,2,3\}$, i.e., no user drops over the first round, the server must recover $W_1 + W_2 + W_3$ when any $K-U = 1$ user drops over the second round. From any two second round messages $Y_{k_1}^{\{1,2,3\}}, Y_{k_2}^{\{1,2,3\}}, k_1, k_2 \in \{1,2,3\}, k_1 \neq k_2$, the server can recover $S_1 + S_2 + S_3$, due to the assignment using an $\mbox{MDS}_{3\times 2}$ matrix (see (\ref{eq:mds})). Then from the first round messages, the server can have $X_1 + X_2 + X_3 = W_1 + W_2 + W_3 + S_1 + S_2 + S_3$. Equipped with $S_1 + S_2 + S_3$, the desired sum $W_1 + W_2 + W_3$ can be decoded with no error.

{\em Security:} The intuition of the security of the achievable scheme is that the first round messages are protected by independent randomness variables and the second round messages just give merely sufficient randomness information (and no more) to unlock the desired sum. We verify that the security constraint (\ref{sec}) is satisfied.

When $\mathcal{U}_1 = \{1,2\}$, we have
\begin{eqnarray}
&& I\left( W_1, W_2, W_3;  X_1, X_2, X_3, Y_1^{\{1,2\}}, Y_2^{\{1,2\}} \Big| W_1 + W_2 \right) \notag \\
&=& H\left( X_1, X_2, X_3, Y_1^{\{1,2\}}, Y_2^{\{1,2\}} \Big| W_1 + W_2 \right) - H\left( X_1, X_2, X_3, Y_1^{\{1,2\}}, Y_2^{\{1,2\}} \Big| W_1, W_2, W_3\right)\\
&=& H(W_1 + S_1, W_2 + S_2, W_3 + S_3, S_1 + S_2 | W_1 + W_2) - H(S_1, S_2, S_3 | W_1, W_2, W_3) \label{eq:p1}\\
&\overset{(\ref{h1})}{=}& H(W_1 + S_1, W_2 + S_2, W_3 + S_3 | W_1 + W_2) - H(S_1, S_2, S_3) \label{eq:pp} \\
&\leq& 6 - 6 = 0
\end{eqnarray}
where in (\ref{eq:p1}) we plug in the design of the randomness and message variables, and in (\ref{eq:pp}) the first term follows from the fact that $S_1 + S_2$ can be obtained from $W_1 + S_1, W_2 + S_2, W_1+W_2$.
In the last step, for the first term we use the fact that $W_1 + S_1, W_2 + S_2, W_3 + S_3$ contains at most $6$ symbols from $\mathbb{F}_q$ and uniform distribution maximizes entropy. As mutual information is non-negative, it must be exactly zero when it is smaller than or equal to zero.
The security proof for other cases where $|\mathcal{U}_1| = 2$ follows similarly.

When $\mathcal{U}_1 = \{1,2,3\}$, we have
\begin{eqnarray}
&& I\left( W_1, W_2, W_3;  X_1, X_2, X_3, Y_1^{\{1,2,3\}}, Y_2^{\{1,2,3\}}, Y_3^{\{1,2,3\}} \Big| W_1 + W_2 + W_3 \right) \notag \\
&=& H\left(W_1 + S_1, W_2 + S_2, W_3 + S_3, S_1 + S_2 + S_3 | W_1 + W_2 + W_3\right) \notag\\
&&~- H(S_1, S_2, S_3 | W_1, W_2, W_3) \label{eq:p2}\\
&\overset{(\ref{h1})}{=}& H(W_1 + S_1, W_2 + S_2, W_3 + S_3 | W_1 + W_2 + W_3) - H(S_1, S_2, S_3) \\
&\leq& 6 - 6 = 0
\end{eqnarray}
where in (\ref{eq:p2}), we use the fact that $Y_1^{\{1,2,3\}}, Y_2^{\{1,2,3\}}, Y_3^{\{1,2,3\}}$ is invertible to $S_1 + S_2 + S_3$.

{\em Rate:} As the first round message contains $L_X = 2$ symbols each and the second round message contains $L_Y = 1$ symbol each, the rate achieved is $R_1 = L_X/L = 1$ and $R_2 = L_Y/L = 1/2$, which matches Theorem \ref{thm:main}. 

\subsection{Example 2: $K = 3, U = 2, T = 1$}
Continuing from the above example, we increase $T$ from $T=0$ to $T=1$, i.e., the server could collude with any single user. The new element needed here is to inject additional noise in sharing the sum of randomness variables used in the first round messages. Note that while this coding idea is simple to describe,  the security proof becomes more involved.

Suppose $L = U-T = 1$, i.e., $W_k \in \mathbb{F}_q$ and $q \geq 5$. The achievable scheme is described as follows.

{\em Randomness Assignment:}
Consider $7$ i.i.d. uniform symbols over $\mathbb{F}_q$, denoted as $S_1, S_2, S_3$, $N_1, N_2, N_3, N_4$ and yield the following generic linear combinations of the sum of some subsets of $\{S_1, S_2, S_3\}$ and some additional noise variable $N_i$.
\begin{eqnarray}
~ \left[ \begin{array}{c}
Z_1^{\{1,2\}}\\
Z_2^{\{1,2\}}
\end{array}
\right] \triangleq {\bf C}_{2\times 2} \left[ \begin{array}{c}
S_1 + S_2\\
N_1
\end{array}
\right], &&
\left[ \begin{array}{c}
Z_1^{\{1,3\}}\\
Z_3^{\{1,3\}}
\end{array}
\right] \triangleq {\bf C}_{2\times 2} \left[ \begin{array}{c}
S_1 + S_3 \\
N_2
\end{array}
\right], \notag \\
~ \left[ \begin{array}{c}
Z_2^{\{2,3\}}\\
Z_3^{\{2,3\}}
\end{array}
\right] \triangleq {\bf C}_{2\times 2} \left[ \begin{array}{c}
S_2 + S_3 \\
N_3
\end{array}
\right], &&
\left[ \begin{array}{c}
Z_1^{\{1,2,3\}}\\
Z_2^{\{1,2,3\}}\\
Z_3^{\{1,2,3\}}
\end{array}
\right] \triangleq {\bf C}_{3 \times 2} \left[ \begin{array}{c}
S_1 + S_2 + S_3\\
N_4
\end{array}
\right] \label{eq:mds2}
\end{eqnarray}
where ${\bf C}_{a\times b}$ denotes a Cauchy matrix of dimension $a \times b$, i.e., the element in the $i$-th row and $j$-column is set as 
\begin{eqnarray}
c_{ij} = \frac{1}{\alpha_i - \beta_j}, ~\alpha_i, \beta_j, i \in [a], j \in [b] ~\mbox{are distinct over $\mathbb{F}_q$.}~
\end{eqnarray}
Note that $q \geq 5$, so distinct elements as required above exist over $\mathbb{F}_q$. Intuitively, Cauchy matrices are used to ensure that the independent noise variables $N_i$ are fully mixed with the sum of $S_k$ variables to avoid any unwanted leakage (see the proof below). Then we set
\begin{eqnarray}
Z_k &=& \left(S_k,  \left(Z_k^{\mathcal{U}_1} \right)_{\mathcal{U}_1: k \in \mathcal{U}_1 \subset \{1,2,3\}, |\mathcal{U}_1|\geq 2} \right), \forall k \in \{1,2,3\}. \label{eq:z}
\end{eqnarray}

{\em Message Generation:} For the first round, we set
\begin{eqnarray}
X_1 = W_1 + S_1, ~X_2 = W_2 + S_2, ~X_3 = W_3 + S_3.
\end{eqnarray}
For the second round, we set
\begin{eqnarray}
\forall \mathcal{U}_1 \subset \{1,2,3\}, |\mathcal{U}_1| \geq 2 : ~ Y_k^{\mathcal{U}_1} = Z_k^{\mathcal{U}_1}, ~\forall k \in \mathcal{U}_1. \label{eq:ans}
\end{eqnarray}

{\em Proof of Correctness:} 
For any $\mathcal{U}_1$ such that $|\mathcal{U}_1| \geq 2$, due to the randomness and message design (see (\ref{eq:mds2}) and (\ref{eq:ans})), the server can recover $\sum_{k\in\mathcal{U}_1} S_k$ from any set of second round messages where $|\mathcal{U}_2| \geq 2$. Then from $\sum_{k\in\mathcal{U}_1} X_k = \sum_{k\in\mathcal{U}_1} W_k + \sum_{k\in\mathcal{U}_1} S_k$, the server can decode the desired sum aggregation $\sum_{k\in\mathcal{U}_1} W_k$ with no error.

{\em Proof of Security:} We show that the injected noise variables $N_i$ help to guarantee the security constraint (\ref{sec}) under collusion.

Suppose $\mathcal{U}_1 = \{1,2\}$ and the colluding user set is $\mathcal{T} = \{1\}$, then we have
\begin{eqnarray}
&& I\left( W_1, W_2, W_3;  X_1, X_2, X_3, Y_1^{\{1,2\}}, Y_2^{\{1,2\}} \Big| W_1 + W_2, W_1, Z_1 \right) \notag \\
&=& H\left( X_1, X_2, X_3, Y_1^{\{1,2\}}, Y_2^{\{1,2\}} \Big| W_1 + W_2, W_1, Z_1 \right) \notag\\
&&~- H\left( X_1, X_2, X_3, Y_1^{\{1,2\}}, Y_2^{\{1,2\}} \Big| W_1, W_2, W_3, Z_1\right) \label{eq:inv}\\
&=& H(W_1 + S_1, W_2 + S_2, W_3 + S_3, S_1 + S_2, N_1 | W_1 + W_2, W_1, Z_1) \notag\\
&&~- H(S_1, S_2, S_3, N_1 | W_1, W_2, W_3, Z_1) \\
&\overset{(\ref{h1})}{=}& H(S_1, W_2 + S_2, W_3 + S_3, N_1 | W_1 + W_2, W_1, Z_1) - H(S_1, S_2, S_3, N_1 | Z_1) \\
&=& H(W_2+S_2, W_3 + S_3 | W_1+W_2, W_1, Z_1) + \underbrace{H(N_1| W_2+S_2, W_3 + S_3, W_1+W_2, W_1, Z_1)}_{=0} \notag\\
&&~- H(S_2, S_3 | Z_1) - \underbrace{H(N_1 | Z_1, S_2, S_3)}_{=0} \label{eq:can}\\
&\leq& 2 - 2 = 0
\end{eqnarray}
where (\ref{eq:can}) is due to the fact that $S_1$ is contained in $Z_1$ (see (\ref{eq:z})) and $N_1$ can be obtained from $Z_1^{\{1,2\}}$ (contained in $Z_1$), when $S_1 + S_2$ is known (obtained from $Z_1, S_2$, refer to (\ref{eq:mds2}), (\ref{eq:z})).
In the last step, the first term follows from the property that uniform random variables are entropy maximizers, and the second term is due to the independence of $(S_2, S_3)$ and $Z_1$, whose proof will be presented in Lemma \ref{lemma:indp} when we give the general proof.

The security proof for other cases of $\mathcal{U}_1$ and $\mathcal{T}$ is similar to that above, which is omitted here and deferred to the general proof presented in the next section.

{\em Rate Calculation:} As $L_X = L_Y = 1$ symbol, we have $R_1 = R_2 = 1$, as desired for this case. 

\subsection{General Proof for Arbitrary $K, U, T$}
The achievability proof for arbitrary $K, U, T$ is an immediate generalization of that of the above two examples. We first consider the case where the field size $q$ is no smaller than $K+U$, and then show that the proof can be adapted with a minor change to cover all other field sizes. As $\mathcal{R}^* = \emptyset$ when $U \leq T$, we only need to consider settings where $U > T$.

\subsubsection*{Large fields: $q \geq K+U$}
Suppose $L = U-T$, i.e., $W_k \in \mathbb{F}_q^{L\times 1}$ and $q \geq K+U$. 

{\em Randomness Assignment:}
Consider $K$ i.i.d. uniform $L \times 1$ vectors over $\mathbb{F}_q$, denoted as $S_k, \forall k \in [K]$. Consider $\sum_{u = U}^{K} \binom{K}{u}$ i.i.d. uniform $T \times 1$ vectors over $\mathbb{F}_q$, denoted as $N^{\mathcal{U}_1}, \forall \mathcal{U}_1 \subset [K], |\mathcal{U}_1| \geq U$. $(S_k)_k$ and $(N^{\mathcal{U}_1})_{\mathcal{U}_1}$ are independent.
Then yield generic linear combinations of the sum of some $S_k$ variables and some $N^{\mathcal{U}_1}$ variable as follows. For any $\mathcal{U}_1$ such that $\mathcal{U}_1 \subset [K], |\mathcal{U}_1| \geq U$, we set\footnote{The design of the $Z_k^{\mathcal{U}_1}$ variables guarantees that any collection of at most $T$ variables in $Z_k^{\mathcal{U}_1}$ reveal nothing about the secret $\sum_{k\in\mathcal{U}_1} S_k$ and any collection of at least $U$ variables in $Z_k^{\mathcal{U}_1}$ can recover the secret. This design is essentially a threshold {\em ramp} secret sharing scheme \cite{Blakley_Meadows, Yamamoto, Kurihara_Kiyomoto_Fukushima_Tanaka}, which is a generalization of classic threshold secret sharing \cite{Shamir, Blakley, McEliece_Sarwate}.}
\begin{eqnarray}
\left( Z_k^{\mathcal{U}_1} \right)_{k \in \mathcal{U}_1} 
= {\bf C}_{|\mathcal{U}_1| \times (L+T)} \left[ \begin{array}{c}
\sum_{k\in\mathcal{U}_1} S_k \vspace{0.05in}\\
N^{\mathcal{U}_1}
\end{array}
\right] = {\bf C}_{|\mathcal{U}_1| \times U} \left[ \begin{array}{c}
\sum_{k\in\mathcal{U}_1} S_k \vspace{0.05in}\\
N^{\mathcal{U}_1}
\end{array}
\right] \label{eq:mds3}
\end{eqnarray}
where ${\bf C}_{a\times b}$ denotes a Cauchy matrix of dimension $a \times b$, i.e., the element in the $i$-th row and $j$-column is $c_{ij} = \frac{1}{\alpha_i - \beta_j}$, where $\alpha_i, \beta_j, i \in [a], j \in [b] ~\mbox{are distinct over $\mathbb{F}_q$.}$ Note that $|\mathcal{U}_1| + U \leq K + U \leq q$, so the required distinct elements exist over $\mathbb{F}_q$. 
Then we set
\begin{eqnarray}
Z_k &=& \left(S_k,  \left(Z_{k}^{\mathcal{U}_1} \right)_{\mathcal{U}_1: k \in \mathcal{U}_1 \subset [K], |\mathcal{U}_1|\geq U} \right), \forall k \in [K]. \label{eq:zz}
\end{eqnarray}

To prepare for the security proof, we present some useful properties on the entropy of the randomness variables in the following lemma.
\begin{lemma}\label{lemma:indp}
For the random variables defined above, for any $\mathcal{U}_1 \subset [K], |\mathcal{U}_1| \geq U$, any $\mathcal{T} \subset [K], |\mathcal{T}| \leq T$, and any $\mathcal{T}' \subset [K], \mathcal{T}' \cap \mathcal{T} = \emptyset$, we have
\begin{itemize}
\item $\left(Z_k \right)_{k\in\mathcal{T}}$ is uniform and is independent of $\left(S_k\right)_{k\in\mathcal{T}'}$.
\begin{eqnarray}
&& H\left( \left( Z_k \right)_{k\in\mathcal{T}} \right) = H\left(\left( Z_k \right)_{k\in\mathcal{T}} \big| \left(S_k\right)_{k\in\mathcal{T}'} \right) = |\mathcal{T}| \left(L + \sum_{u = U-1}^{K-1} \binom{K-1}{u} \right), \label{lemma:e1}\\
&& H\left(\left(S_k\right)_{k\in\mathcal{T}'} \big| \left( Z_k \right)_{k\in\mathcal{T}} \right) = |\mathcal{T}'| L. \label{lemma:e2}
\end{eqnarray}
\item Given either $\sum_{k\in\mathcal{U}_1} S_k$ or $\left(S_k\right)_{k\in[K]}$, $\left( Z_k \right)_{k\in\mathcal{T}}$ contains $|\mathcal{T} \cap \mathcal{U}_1|$ linearly independent combinations of the $T$ i.i.d. symbols in $N^{\mathcal{U}_1}$. 
\begin{eqnarray}
H\left( N^{\mathcal{U}_1} \Bigg| \sum_{k\in\mathcal{U}_1} S_k, \left( Z_k \right)_{k\in\mathcal{T}} \right) = 
H\left(N^{\mathcal{U}_1} \Bigg| \left(S_k\right)_{k\in[K]}, \left( Z_k \right)_{k\in\mathcal{T}} \right) =
T - |\mathcal{T} \cap \mathcal{U}_1|. \label{lemma:e3}
\end{eqnarray}
\end{itemize}
\end{lemma}

The detailed proof of Lemma \ref{lemma:indp} is deferred to Section \ref{sec:lemma_indp}.

{\em Message Generation:} We set
\begin{eqnarray}
&& X_k = W_k + S_k, \forall k \in [K], \notag \\
&& \forall \mathcal{U}_1 \subset[K], |\mathcal{U}_1| \geq U: ~Y_k^{\mathcal{U}_1} = Z_k^{\mathcal{U}_1}, ~\forall k \in \mathcal{U}_1. \label{eq:anss}
\end{eqnarray}

{\em Proof of Correctness:} 
For any $\mathcal{U}_1$ such that $|\mathcal{U}_1| \geq U$, as any square sub-matrix of a Cauchy matrix (with distinct $\alpha_i, \beta_j$) has full rank \cite{Schechter}, the server can recover $\sum_{k\in\mathcal{U}_1} S_k$ from any $U$ second round messages. Combining with the first round messages, the server can have $\sum_{k\in\mathcal{U}_1} X_k = \sum_{k\in\mathcal{U}_1} W_k + \sum_{k\in\mathcal{U}_1} S_k$ and then decode the desired sum aggregation $\sum_{k\in\mathcal{U}_1} W_k$ with no error.

{\em Proof of Security:} Consider any $\mathcal{U}_1 \subset [K], |\mathcal{U}_1| \geq U$ and any $\mathcal{T} \subset [K], |\mathcal{T}| \leq T$. We verify that the security constraint (\ref{sec}) is satisfied. Denote the difference of two sets $\mathcal{A}, \mathcal{B}$ as $\mathcal{A}\backslash\mathcal{B}$, i.e., the set of elements that belong to $\mathcal{A}$ but not $\mathcal{B}$.
\begin{eqnarray}
&& I\left(\left(W_k\right)_{k\in[K]}; \left(X_k\right)_{k\in[K]}, \left(Y_k^{\mathcal{U}_1}\right)_{k\in\mathcal{U}_1} \Bigg| \sum_{k\in\mathcal{U}_1} W_k, \left( W_k, Z_k \right)_{k\in\mathcal{T}} \right) \notag\\
&=& H\left( \left(X_k\right)_{k\in[K]}, \left(Y_k^{\mathcal{U}_1}\right)_{k\in\mathcal{U}_1} \Bigg| \sum_{k\in\mathcal{U}_1} W_k, \left( W_k, Z_k \right)_{k\in\mathcal{T}} \right) \notag\\
&& ~- H\left(\left(X_k\right)_{k\in[K]}, \left(Y_k^{\mathcal{U}_1}\right)_{k\in\mathcal{U}_1} \Bigg| \left(W_k\right)_{k\in[K]}, \left( Z_k \right)_{k\in\mathcal{T}} \right) \\
&=& H\left( \left(W_k + S_k\right)_{k\in[K]}, \sum_{k\in\mathcal{U}_1} S_k, N^{\mathcal{U}_1} \Bigg| \sum_{k\in\mathcal{U}_1} W_k, \left( W_k, Z_k \right)_{k\in\mathcal{T}} \right) \notag\\
&& ~- H\left(\left(S_k\right)_{k\in[K]}, \sum_{k\in\mathcal{U}_1} S_k, N^{\mathcal{U}_1} \Bigg| \left(W_k\right)_{k\in[K]}, \left( Z_k \right)_{k\in\mathcal{T}} \right) \label{eq:e1} \\
&\leq& H\left( \left(W_k + S_k\right)_{k\in[K]\backslash\mathcal{T}} \Bigg| \sum_{k\in\mathcal{U}_1} W_k, \left( W_k, Z_k \right)_{k\in\mathcal{T}} \right) + H\left( N^{\mathcal{U}_1} \Bigg| \sum_{k\in\mathcal{U}_1} S_k, \left( Z_k \right)_{k\in\mathcal{T}} \right) \notag\\ 
&& ~- H\left(\left(S_k\right)_{k\in[K]\backslash\mathcal{T}} \Bigg| \left( Z_k \right)_{k\in\mathcal{T}} \right)
- H\left(N^{\mathcal{U}_1} \Bigg| \left(S_k\right)_{k\in[K]}, \left( Z_k \right)_{k\in\mathcal{T}} \right) \label{eq:e2} \\
&\leq& (K - |\mathcal{T}|)L + (T - |\mathcal{T} \cap \mathcal{U}_1|)  - (K - |\mathcal{T}|)L -( T - |\mathcal{T} \cap \mathcal{U}_1| ) = 0
\end{eqnarray}
where in (\ref{eq:e1}) we plug in the design of the message variables (\ref{eq:anss}) and use the fact that $(Y_k^{\mathcal{U}_1})_{k\in\mathcal{U}_1}$ is invertible to $(\sum_{k\in\mathcal{U}_1} S_k, N^{\mathcal{U}_1})$ (see (\ref{eq:mds3}), (\ref{eq:anss})), and in (\ref{eq:e2}) we use the chain rule and the independence of the inputs $W_k$ and the randomness variables $Z_k$. In the last step, the first term follows from the fact that uniform variables maximize entropy, and other terms follow from Lemma \ref{lemma:indp}.

{\em Rate Calculation:} As $L_X = L, L_Y = 1$, we have $R_1 = 1, R_2 = 1/L = 1/(U-T)$, as desired. 
The achievability proof is thus completed.

\begin{remark}\label{remark:noind}
We can verify that both the correctness proof and the security proof do not use the independence and uniformity of the input vectors $W_1, \cdots, W_K$. As such, the rate tuple $(R_1, R_2) = (1, \frac{1}{U-T})$ is achievable for arbitrarily distributed inputs $W_1, \cdots, W_K$.
\end{remark}


\subsubsection*{Any field size}
We consider an arbitrary field $\mathbb{F}_q$, where $q = p^n$ for a prime $p \geq 2$ and an integer $n \geq 1$. The proof for the above $q \geq K+U$ case only relies on the property that the field size is sufficiently large so that there exist a required number of distinct elements. Here for arbitrary field size, we `amplify' the field size by grouping a number of field elements (say $B$ elements) from $\mathbb{F}_{p^n}$ and view such a group of elements as one element from the extension field $\mathbb{F}_{p^{Bn}}$. That is, we set $L = B\overline{L}$ so that
\begin{eqnarray}
W_k = (W_k(l))_{l \in [B\overline{L}]} = (\overline{W}_k(l))_{l\in[\overline{L}]} , \forall k \in [K]
\end{eqnarray}
where $\overline{W}_k(l) \triangleq \big(W_k((l-1)B + 1); W_k((l-1)B + 2); \cdots; W_k(lB) \big) \in \mathbb{F}_{p^n}^{B \times 1}, \forall l \in [\overline{L}]$. Then the desired element-wise vector sum can be performed equivalently as element-wise vector sum over the extension field,
\begin{eqnarray}
\left( \sum_{k\in\mathcal{U}_1} W_k \right)_{\mbox{\footnotesize sum over $\mathbb{F}_{p^n}$}} \Longleftrightarrow \left( \sum_{k\in\mathcal{U}_1} \overline{W}_k \right)_{\mbox{\footnotesize sum over $\mathbb{F}_{p^{Bn}}$}}.
\end{eqnarray}
Thus, by grouping elements (i.e., increasing the input length by a factor of $B$), we have created an equivalent problem over a larger field (i.e., the field size is increased by a power of $B$). We can now apply the same coding scheme above, which will work as long as the extended field size $q^B$ satisfies that $q^B \geq K+U$ (as required by the scheme above). Such an extension is possible because in Shannon theoretic formulation, the input length $L$ is allowed to approach infinity. Specifically, over any field $\mathbb{F}_q$, $L$ can be set to $B (U-T) = \lceil \log_q(K+U) \rceil (U-T)$, where $U-T$ is the required input length for the scheme above. Therefore, the same rate tuple $(R_1, R_2) = (1, 1/(U-T))$ can be achieved over any finite field.

\subsection{Proof of Lemma \ref{lemma:indp}}\label{sec:lemma_indp}
First, consider (\ref{lemma:e1}). Note that $\left(Z_k \right)_{k\in\mathcal{T}}$ contains $|\mathcal{T}| \left(L + \sum_{u = U-1}^{K-1} \binom{K-1}{u} \right)$ symbols from $\mathbb{F}_q$. As uniform variables maximize entropy and conditioning cannot increase entropy, it suffices to prove
\begin{eqnarray}
H\left(\left( Z_k \right)_{k\in\mathcal{T}} \big| \left(S_k\right)_{k\in\mathcal{T}'} \right) \geq |\mathcal{T}| \left(L + \sum_{u = U-1}^{K-1} \binom{K-1}{u} \right)
\end{eqnarray}
and this proof is presented next.
\begin{eqnarray}
&& H\left(\left( Z_k \right)_{k\in\mathcal{T}} \big| \left(S_k\right)_{k\in\mathcal{T}'} \right) \notag\\
&\overset{(\ref{eq:zz})}{=}& H\left(\left( S_{k},  \left(Z_{k}^{\mathcal{U}_1} \right)_{\mathcal{U}_1: k \in \mathcal{U}_1 \subset [K], |\mathcal{U}_1|\geq U}  \right)_{k\in\mathcal{T}} \bigg| \left(S_k\right)_{k\in\mathcal{T}'} \right) \\
&=& H\left(\left( S_k \right)_{k\in\mathcal{T}} \big| \left(S_k\right)_{k\in\mathcal{T}'}\right) + H\left(\left( \left(Z_{k}^{\mathcal{U}_1} \right)_{\mathcal{U}_1: k \in \mathcal{U}_1 \subset [K], |\mathcal{U}_1|\geq U}  \right)_{k\in\mathcal{T}} \bigg| \left(S_k\right)_{k\in\mathcal{T}\cup \mathcal{T}'} \right) \\
&\geq& |\mathcal{T}|L + H\left(\left( \left(Z_{k}^{\mathcal{U}_1} \right)_{k\in\mathcal{T}}  \right)_{\mathcal{U}_1: \mathcal{T} \cap \mathcal{U}_1 \neq \emptyset, \mathcal{U}_1 \subset [K], |\mathcal{U}_1|\geq U} \bigg| \left(S_k\right)_{k\in[K]} \right) \label{eq:exp1} \\
&=&  |\mathcal{T}|L + \sum_{{\mathcal{U}_1: \mathcal{T} \cap \mathcal{U}_1 \neq \emptyset, \mathcal{U}_1 \subset [K], |\mathcal{U}_1|\geq U}} H\left( {\bf C}^{sub}_{ |\mathcal{T} \cap \mathcal{U}_1| \times T} \times N^{\mathcal{U}_1}   \right) \label{eq:exp3}\\
&=&  |\mathcal{T}|L + \sum_{{\mathcal{U}_1: \mathcal{T} \cap \mathcal{U}_1 \neq \emptyset, \mathcal{U}_1 \subset [K], |\mathcal{U}_1|\geq U}}  |\mathcal{T} \cap \mathcal{U}_1|  \label{eq:exp4} \\
&=& |\mathcal{T}|L + \sum_{t \in \mathcal{T}} \sum_{{\mathcal{U}_1: t \in \mathcal{U}_1, \mathcal{U}_1 \subset [K], |\mathcal{U}_1|\geq U}}  1 =   |\mathcal{T}|L +  |\mathcal{T}| \sum_{u=U-1}^{K-1} \binom{K-1}{u}
\end{eqnarray}
where in (\ref{eq:exp1}), the first term follows from the fact that the $S_k$ variables are independent, and the second term switches the order of $k$ and $\mathcal{U}_1$ in the counting. 
In (\ref{eq:exp3}), the second term is due to the properties that 1) $(S_k)_k$ and $(N^{\mathcal{U}_1})_{\mathcal{U}_1}$ are independent, and 2) conditioned on all $S_k$ variables, $(Z_{k}^{\mathcal{U}_1})_{k\in\mathcal{T}}$ is invertible to  a number of linear combinations of the symbols in $N^{\mathcal{U}_1}$, with coefficients given by a sub-matrix of the Cauchy matrix ${\bf C}_{|\mathcal{U}_1| \times U}$, denoted as ${\bf C}^{sub}_{}$. In (\ref{eq:exp4}), the second term follows from the fact that the sub-matrix of a Cauchy matrix has full rank \cite{Schechter} and $N^{\mathcal{U}_1}$ consists of i.i.d. uniform symbols. In the last step, we count through each element of the set $\mathcal{T}$.

Second, consider (\ref{lemma:e2}), which is an immediate consequence of (\ref{lemma:e1}).
\begin{eqnarray}
 H\left(\left(S_k\right)_{k\in\mathcal{T}'} \big| \left( Z_k \right)_{k\in\mathcal{T}} \right) 
&=& H\left(\left(S_k\right)_{k\in\mathcal{T}'} \right) - I\left(\left(S_k\right)_{k\in\mathcal{T}'} ; \left( Z_k \right)_{k\in\mathcal{T}} \right) \\
&\overset{(\ref{lemma:e1})}{=}& |\mathcal{T}'| L .
\end{eqnarray}

Third, consider (\ref{lemma:e3}). It suffices to prove
\begin{eqnarray}
T - |\mathcal{T} \cap \mathcal{U}_1| \geq H\left( N^{\mathcal{U}_1} \Bigg| \sum_{k\in\mathcal{U}_1} S_k, \left( Z_k \right)_{k\in\mathcal{T}} \right) \geq
H\left(N^{\mathcal{U}_1} \Bigg| \left(S_k\right)_{k\in[K]}, \left( Z_k \right)_{k\in\mathcal{T}} \right) \geq 
T - |\mathcal{T} \cap \mathcal{U}_1|.
\end{eqnarray}
The middle `$\geq$' follows from the fact that conditioning cannot increase entropy and we are left to prove the two remaining `$\geq$'. To prove the left `$\geq$', note that from $(\sum_{k\in\mathcal{U}_1} S_k, \left( Z_k \right)_{k\in\mathcal{T}})$, we obtain $|\mathcal{T} \cap \mathcal{U}_1|$ linearly independent combinations of the symbols in $N^{\mathcal{U}_1}$ (see (\ref{eq:mds3})). Then we have
\begin{eqnarray}
H\left( N^{\mathcal{U}_1} \Bigg| \sum_{k\in\mathcal{U}_1} S_k, \left( Z_k \right)_{k\in\mathcal{T}} \right) 
&=& H\left( N^{\mathcal{U}_1} \Bigg| \sum_{k\in\mathcal{U}_1} S_k, \left( Z_k \right)_{k\in\mathcal{T}},  {\bf C}^{sub}_{|\mathcal{T} \cap \mathcal{U}_1| \times T} \times N^{\mathcal{U}_1}\right) \\
&\leq& H\left( N^{\mathcal{U}_1} \Bigg| {\bf C}^{sub}_{|\mathcal{T} \cap \mathcal{U}_1| \times T} \times N^{\mathcal{U}_1}\right) = T - |\mathcal{T} \cap \mathcal{U}_1|.
\end{eqnarray}
To prove the right `$\geq$', we have
\begin{eqnarray}
&& H\left(N^{\mathcal{U}_1} \Bigg| \left(S_k\right)_{k\in[K]}, \left( Z_k \right)_{k\in\mathcal{T}} \right) \notag\\
&=& H\left(N^{\mathcal{U}_1} \Bigg| \left(S_k\right)_{k\in[K]}, \left( Z_k \right)_{k\in\mathcal{T}}, {\bf C}^{sub}_{|\mathcal{T} \cap \mathcal{U}_1| \times T} \times N^{\mathcal{U}_1} \right)  \\
&\geq& H\left(N^{\mathcal{U}_1} \Bigg| \left(S_k\right)_{k\in[K]}, \left( Z_k \right)_{k\in\mathcal{T}}, {\bf C}^{sub}_{|\mathcal{T} \cap \mathcal{U}_1| \times T} \times N^{\mathcal{U}_1}, \left( N^{\overline{\mathcal{U}}_1} \right)_{ \overline{\mathcal{U}}_1 \neq {\mathcal{U}_1}, \overline{\mathcal{U}}_1 \subset [K], |\overline{\mathcal{U}}_1| \geq U} \right)  \\
&=& H\left( N^{\mathcal{U}_1} \Bigg| {\bf C}^{sub}_{|\mathcal{T} \cap \mathcal{U}_1| \times T} \times N^{\mathcal{U}_1}\right) = T - |\mathcal{T} \cap \mathcal{U}_1|
\end{eqnarray}
where the last line follows from the fact that 1) conditioned on all $(S_k)_k$ and all $(N^{\overline{\mathcal{U}}_1})_{\overline{\mathcal{U}}_1}$ other than $N^{\mathcal{U}_1}$, $\left(Z_k \right)_{k\in\mathcal{T}}$ is left with only symbols from $N^{\mathcal{U}_1}$, and 2) $(S_k)_k$ and $(N^{\overline{\mathcal{U}}_1})_{\overline{\mathcal{U}}_1}$ are independent.

\section{Proof of Theorem \ref{thm:main}: Converse}\label{sec:con}
Let us start with a simple consequence of the independence of inputs $\left(W_k\right)_{k\in[K]}$ and $\left(Z_k\right)_{k\in[K]}$, and the uniformity of $\left(W_k\right)_{k\in[K]}$, which is stated in the following lemma to facilitate later use.

\begin{lemma}\label{lemma:indc}
For any $V_2 < V_1 < K$, the following equality holds. 
\begin{eqnarray}
I \left(\sum_{k\in[V_1]} W_k ; \sum_{k\in[V_1+1]} W_k, \left( W_k, Z_k \right)_{k\in[V_2]} \right) = 0. \label{eq:indc}
\end{eqnarray}
\end{lemma}

{\it Proof of Lemma \ref{lemma:indc}:}
\begin{eqnarray}
&& I \left(\sum_{k\in[V_1]} W_k ; \sum_{k\in[V_1+1]} W_k, \left( W_k, Z_k \right)_{k\in[V_2]} \right) \notag\\
&\overset{(\ref{h1})}{=}& I \left(\sum_{k\in[V_1]} W_k ; \sum_{k\in[V_1+1]} W_k, \left( W_k \right)_{k\in[V_2]} \right) \\
&=&  H \left(\sum_{k\in[V_1]} W_k \right) - H \left(\sum_{k\in[V_1]} W_k \Bigg| \sum_{k\in[V_1+1]} W_k, \left( W_k \right)_{k\in[V_2]} \right) \\
&=& L - H \left( W_{V_1+1} \Bigg| \sum_{k\in[V_1+1]} W_k, \left( W_k \right)_{k\in[V_2]} \right) \label{eq:c1} \\
&=& L - \big( (V_2 +2) L - (V_2 + 1)L \big)= 0
\end{eqnarray}
where in (\ref{eq:c1}) and the last step, we use the uniformity of $\left(W_k\right)_{k\in[K]}$.
\hfill\QED

Next, we present the converse proof for $U \leq T$ and $U > T$ cases (i.e., first and second round message rates), where the key is to judiciously choose $\mathcal{U}_1, \mathcal{U}_2, \mathcal{T}$ to produce the desired bounds from the correctness and security constraints (\ref{corr}), (\ref{sec}).

\subsection{$U \leq T:$ Proof of $\mathcal{R}^* = \emptyset$}
We show that when $U \leq T$, the system constraints are self-contradictory, so they cannot be satisfied by any secure aggregation scheme, i.e., $\mathcal{R}^* = \emptyset$. To see why we have a contradiction, consider $\mathcal{U}_1 = [U+2]$ 
and $
\mathcal{T} = [U]$. Note that $U \leq T \leq K-2$, so this choice of $\mathcal{U}_1$ and $\mathcal{T}$ is feasible. From the security constraint (\ref{sec}), we have
\begin{eqnarray}
0 &=& I\left(\left(W_k\right)_{k\in[K]}; \left(X_k\right)_{k\in[K]}, \left(Y_k^{[U+2]}\right)_{k\in[U+2]} \Bigg| \sum_{k\in[U+2]} W_k, \left( W_k, Z_k \right)_{k\in[U]} \right) \\
&\geq&  I\left( \sum_{k\in[U+1]} W_k; \left(X_k\right)_{k\in[U+1]} \Bigg| \sum_{k\in[U+2]} W_k, \left( W_k, Z_k \right)_{k\in[U]} \right) \\
&\overset{(\ref{2y})}{=}& I\left( \sum_{k\in[U+1]} W_k; \left(X_k\right)_{k\in[U+1]}, \left( Y_k^{[U+1]} \right)_{k\in[U]} \Bigg| \sum_{k\in[U+2]} W_k, \left( W_k, Z_k \right)_{k\in[U]} \right) \label{eq:c2} \\
&=& I\left( \sum_{k\in[U+1]} W_k; \left(X_k\right)_{k\in[U+1]}, \left( Y_k^{[U+1]} \right)_{k\in[U]}, \sum_{k\in[U+2]} W_k, \left( W_k, Z_k \right)_{k\in[U]} \right) \notag\\
&& ~- \underbrace{
I\left( \sum_{k\in[U+1]} W_k; \sum_{k\in[U+2]} W_k, \left( W_k, Z_k \right)_{k\in[U]} \right)}_{= 0} 
\label{eq:c3} \\
&\geq& I\left( \sum_{k\in[U+1]} W_k; \left(X_k\right)_{k\in[U+1]}, \left( Y_k^{[U+1]} \right)_{k\in[U]} \right) \\
&=& H\left( \sum_{k\in[U+1]} W_k \right) -  H\left( \sum_{k\in[U+1]} W_k \Bigg| \left(X_k\right)_{k\in[U+1]}, \left( Y_k^{[U+1]} \right)_{k\in[U]} \right)\label{eq:c4} \\
&\overset{(\ref{corr})}{=}& L - 0 = L\\
\Rightarrow ~0 &\geq& L
\end{eqnarray}
where in (\ref{eq:c2}), we use the fact that $Y_k^{[U+1]}$ is a function of $W_k, Z_k$ (see (\ref{2y})); note that here the choice of the superscript of $Y_k^{[U+1]}$ is crucial (i.e., the first round responding user set). The second term of (\ref{eq:c3}) is zero because of Lemma \ref{lemma:indc}, where we set $V_1 = U+1, V_2 = U$ in (\ref{eq:indc}). In (\ref{eq:c4}), the first term is $L$ because the inputs are independent and uniform so that the sum is also uniform; the second term is zero because of the correctness constraint (\ref{corr}), when $\mathcal{U}_1 = [U+1]$ and $\mathcal{U}_2 = [U]$. In the final step, where $0 \geq L$, we arrive at a contradiction, i.e., the constraints used in the above derivation cannot hold simultaneously. The proof of $\mathcal{R}^* = \emptyset$ is thus complete.

\begin{remark}
The intuition of the above proof is as follows. When $\mathcal{U}_1 = [U+2], \mathcal{T} = [U]$, the security constraint requires that nothing beyond $W_{U+1} + W_{U+2}$ shall be learned, given all the messages and the information from colluding users. However, such messages and colluding information can fully recover all responding messages when $\mathcal{U}_1 = [U+1], \mathcal{U}_2 = [U]$, so from the correctness constraint, $\sum_{k\in[U+1]}W_{k}$ can be decoded and then $W_{U+1}$ can be obtained (given the colluding information), which violates that only $W_{U+1} + W_{U+2}$ shall be learned. The above proof formalizes this intuition.
\end{remark}

\subsection{$U > T:$ Proof of $R_1 \geq 1$}
We prove the converse bound for the first round message rate. Consider any $u \in [K]$, and set $\mathcal{U}_1 = [K], \mathcal{U}_2 = [K]\backslash\{u\}$. Then from the correctness constraint (\ref{corr}), we have
\begin{eqnarray}
0 &=& H\left(\sum_{k\in[K]} W_k \Bigg| \left(X_k\right)_{k\in[K]}, \left(Y_k^{[K]}\right)_{k\in[K]\backslash\{u\}} \right) \\
&\geq& H\left(\sum_{k\in[K]} W_k \Bigg| \left(X_k\right)_{k\in[K]}, \left(Y_k^{[K]}\right)_{k\in[K]\backslash\{u\}}, \left( W_k, Z_k \right)_{k\in[K]\backslash\{u\}} \right) \\
&\overset{(\ref{1x})(\ref{2y})}{=}& H\left( W_{u} \big| X_{u}, \left( W_k, Z_k \right)_{k\in[K]\backslash\{u\}} \right) \label{eq:cc1}
\end{eqnarray}
where (\ref{eq:cc1}) follows from the fact that $ \left(X_k \right)_{k\in[K]\backslash\{u\}}, (Y_k^{[K]})_{k\in[K]\backslash\{u\}}$ are functions of $\left( W_k, Z_k \right)_{k\in[K]\backslash\{u\}}$ (see (\ref{1x}), (\ref{2y})).
Next,
\begin{eqnarray}
L &\overset{(\ref{h2})}{=}& H(W_{u}) \overset{(\ref{h1})}{=} H\left( W_{u} \big|  \left( W_k, Z_k \right)_{k\in[K]\backslash\{u\}} \right) \\
&\overset{(\ref{eq:cc1})}{=}& I\left( W_{u} ; X_{u} \big| \left( W_k, Z_k \right)_{k\in[K]\backslash\{u\}} \right) \\
&\leq& H\left(X_{u} \big|  \left( W_k, Z_k \right)_{k\in[K]\backslash\{u\}} \right) \label{eq:cr1} \\ 
&\leq& H\left(X_{u} \right) ~\leq~ L_X \label{eq:cr2}\\
\Rightarrow~ R_1 \overset{(\ref{rate})}{=} \frac{L_X}{L} &\geq& 1.
\end{eqnarray}


\begin{remark}
Intuitively, the reason that the first round message length $L_X$ shall be no smaller than the input size $L$ is as follows. Note that any user, say User $u$, may survive over the first round, but may drop over the second round. Also, User $u$'s input is only available at User $u$. As a result, the first round message from User $u$ (which has length $L_X$) must at least contain all information contained in Input $u$, $W_u$ (whose entropy is $L$), so that the server may decode a sum function that includes $W_u$. This explanation translates to the proof presented above.
\end{remark}

\subsection{$U > T:$ Proof of $R_2 \geq \frac{1}{U-T}$}
We prove the converse bound for the second round message rate. Consider $\mathcal{U}_1 = [U+1], \mathcal{U}_2 = [U], \mathcal{T} = [T]$. Then from the security constraint (\ref{sec}), we have
\begin{eqnarray}
0 &=& I\left(\left(W_k\right)_{k\in[K]}; \left(X_k\right)_{k\in[K]}, \left(Y_k^{[U+1]}\right)_{k\in[U+1]} \Bigg| \sum_{k\in[U+1]} W_k, \left( W_k, Z_k \right)_{k\in[T]} \right) \\
&\geq& I\left( \sum_{k\in[U]} W_k; \left(X_k\right)_{k\in[U]} \Bigg| \sum_{k\in[U+1]} W_k, \left( W_k, Z_k \right)_{k\in[T]} \right) \\
&\overset{(\ref{2y})}{=}& I\left( \sum_{k\in[U]} W_k; \left(X_k\right)_{k\in[U]}, \left(Y_k^{[U]}\right)_{k\in[T]} \Bigg| \sum_{k\in[U+1]} W_k, \left( W_k, Z_k \right)_{k\in[T]} \right) \\
&=&  I\left( \sum_{k\in[U]} W_k; \left(X_k\right)_{k\in[U]}, \left(Y_k^{[U]}\right)_{k\in[T]}, \sum_{k\in[U+1]} W_k, \left( W_k, Z_k \right)_{k\in[T]} \right)  \notag\\
&&~- \underbrace{ I\left( \sum_{k\in[U]} W_k; \sum_{k\in[U+1]} W_k, \left( W_k, Z_k \right)_{k\in[T]} \right) }_{\overset{(\ref{eq:indc})}{=} 0} \\
&\geq& I\left( \sum_{k\in[U]} W_k; \left(X_k\right)_{k\in[U]}, \left(Y_k^{[U]}\right)_{k\in[T]} \right). \label{eq:cd}
\end{eqnarray}

Next, consider $\mathcal{U}_1 = \mathcal{U}_2 = [U]$. From the correctness constraint (\ref{corr}), we have
\begin{eqnarray}
0 &=& H\left(\sum_{k\in[U]} W_k \Bigg| \left(X_k\right)_{k\in[U]}, \left(Y_k^{[U]}\right)_{k\in[U]} \right) \\
\Rightarrow~ L &=&  I\left(\sum_{k\in[U]} W_k ; \left(X_k\right)_{k\in[U]}, \left(Y_k^{[U]}\right)_{k\in[U]} \right) \\
&\overset{(\ref{eq:cd})}{=}& I\left(\sum_{k\in[U]} W_k ; \left(Y_k^{[U]}\right)_{k\in[U]\backslash[T]}  \Bigg| \left(X_k\right)_{k\in[U]}, \left(Y_k^{[U]}\right)_{k\in[T]} \right) \\
&\leq& H\left( \left(Y_k^{[U]}\right)_{k\in[U]\backslash[T]}  \right) \leq \sum_{k\in[U]\backslash[T]} H\left( Y_k^{[U]}\right) \leq (U-T) L_Y\\
\Rightarrow~ R_2 &\overset{(\ref{rate})}{=}& \frac{L_Y}{L} \geq \frac{1}{U-T}.
\end{eqnarray}

\begin{remark}
The intuition of the above proof is as follows. Due to the security constraint, all first round messages and any $T$ second round messages do not contribute any useful information in decoding the desired sum (see (\ref{eq:cd})). Then all useful information can only come from the remaining $(U-T)$ second round messages, such that each of them must contain $L/(U-T)$ symbols of information at least on average. Thus the proof, which makes this claim rigorous, follows as above.
\end{remark}

\begin{remark}\label{remark:zn_con}
By checking every step of the converse proof above, we can verify that the converse bounds generalize from the finite field to any group (e.g., including the modulo ring of integers as a special case), where the inputs are uniform over the respective group and `$+$' is replaced by the group operation; additionally, in case of a non-abelian group, the order of `$+$' in $\sum_{k\in\mathcal{U}} W_k$ over the set of elements $(W_k)_{k\in\mathcal{U}}$ is assumed consistently to be increasing in $k$.
\end{remark}

\section{Discussion}\label{sec:dis}
In this section we discuss some interesting observations on aspects beyond the optimal rate characterization of secure aggregation.

\subsection*{Randomness Cost}
Information theoretic security (i.e., statistical independence) is ensured through randomized schemes so that randomness consumption is a meaningful metric and we naturally wish to use as little randomness as possible. The randomness symbols need to be stored at the users to perform the encoding operations, so randomness consumption also translates to storage overhead. While the problem of characterizing the minimum randomness cost of secure aggregation remains open in general, we have obtained some preliminary results on the non-colluding setting, i.e., $T=0$, which are stated in the following theorem.

\begin{theorem}\label{theorem:random}
For the information theoretic secure aggregation problem, with $K$ users, at least $1\leq U \leq K-1$ responding users, and no collusion $(T=0)$, we have
\begin{itemize}
\item{[Total Randomness]}  $\inf_L H\left(\left( Z_k \right)_{k\in[K]} \right)/L = K$ for any $U$;
\item{[Individual Randomness]} when $U = 1$, $\inf_L H(Z_k)/L = K, \forall k \in [K]$.
\end{itemize}
\end{theorem}

The achievability of the total and individual randomness rate stated in Theorem \ref{theorem:random} is proved by the achievable scheme for Theorem \ref{thm:main} (see Section \ref{sec:ach}), where if $T=0$, only $(S_k)_{k\in[K]}$ is used (i.e., $N^{\mathcal{U}_1}$ is an empty vector) and each independent $S_k$ has entropy $L$ so that $H((Z_k)_{k\in[K]}) = KL$; when $U=1$, each $Z_k, \forall k \in [K]$ can recover $(S_k)_{k\in[K]}$ so that $H(Z_k) = KL$. The converse proof is deferred to Section \ref{sec:total} and Section \ref{sec:ind}, and an intuitive explanation is as follows. Due to the security constraint, the $K$ first round messages must be fully protected by independent randomness variables. In addition, from the optimal rate characterization in Theorem \ref{thm:main}, the length of each first round message is at least $L$ so that overall, we need $KL$ randomness symbols at least (just to secure the first round messages). When $U=1$, as we may only see one second round message, each user must hold all $KL$ randomness symbols at hand to ensure correct decoding under all possible choices of first round responding user sets.

When $U > 1$, the minimum individual randomness cost is an open problem even if $T=0$, but we know that the achievable scheme in Theorem \ref{thm:main} can be further optimized. To see this, let us revisit Example 1, where $K=3, U=2, T=0$ and $(H(Z_1)/L, H(Z_2)/L,  H(Z_3)/L) = (2.5, 2.5, 2.5)$ is achieved (see (\ref{eq:zi})). We show that by carefully designing the MDS matrices in (\ref{eq:mds}) (and all other parts of the scheme are not changed), we can achieve $(H(Z_1)/L , H(Z_2)/L , H(Z_3)/L) = (2, 2.5, 2)$. Specifically, we set
\begin{eqnarray}
&& \mbox{MDS}_{2\times 2} = \left[ \begin{array}{cc}
1 & 0\\
0 & 1
\end{array}
\right], \mbox{MDS}_{3\times 2} = \left[ \begin{array}{cc}
1 & 0\\
1 & 1\\
0 & 1
\end{array}
\right] ~\mbox{in (\ref{eq:mds})}~
\notag\\
&\Rightarrow& \left[ \begin{array}{c}
Z_1^{\{1,2\}}\\
Z_2^{\{1,2\}}
\end{array}
\right] = 
\left[ \begin{array}{c}
S_1(1) + S_2(1)\\
S_1(2) + S_2(2)
\end{array}
\right], ~~
\left[ \begin{array}{c}
Z_1^{\{1,3\}}\\
Z_3^{\{1,3\}}
\end{array}
\right] = 
\left[ \begin{array}{c}
S_1(1) + S_3(1)\\
S_1(2) + S_3(2)
\end{array}
\right],
 \notag \\
&& \left[ \begin{array}{c}
Z_2^{\{2.3\}}\\
Z_3^{\{2,3\}}
\end{array}
\right] 
=  \left[ \begin{array}{c}
S_2(1) + S_3(1)\\
S_2(2) + S_3(2)
\end{array}
\right], ~~
\left[ \begin{array}{c}
Z_1^{\{1,2,3\}}\\
Z_2^{\{1,2,3\}}\\
Z_3^{\{1,2,3\}}
\end{array}
\right] =  
\left[ \begin{array}{c}
S_1(1) + S_2(1) + S_3(1)\\
\sum_{l=1}^2 S_1(l) + S_2(l) + S_3(l) \\
S_1(2) + S_2(2) + S_3(2)
\end{array}
\right] \label{eq:newmds}
\end{eqnarray}
where each $S_k(i)$ is i.i.d. and uniform over $\mathbb{F}_q$. Through the above design, we have created certain correlation among $Z_k^{\mathcal{U}_1}$ to reduce $H(Z_k)$, which we now calculate. From (\ref{eq:zi}) and (\ref{eq:newmds}), we have
\begin{eqnarray}
H(Z_1) &=& H\left(S_1, Z_1^{\{1,2\}}, Z_1^{\{1,3\}}, Z_1^{\{1,2,3\}} \right) = H(S_1, S_2(1), S_3(1)) = 4, \notag\\
H(Z_2) &=& H\left(S_2, Z_2^{\{1,2\}}, Z_2^{\{2,3\}}, Z_2^{\{1,2,3\}} \right) = H(S_2, S_1(2), S_3(1), S_1(1) + S_3(2)) = 5, \notag\\
H(Z_3) &=& H\left(S_3, Z_3^{\{1,3\}}, Z_3^{\{2,3\}}, Z_3^{\{1,2,3\}} \right) = H(S_3, S_1(2), S_2(2)) = 4. 
\end{eqnarray}
where $S_k = (S_k(1); S_k(2))$.
As $L=2$, we have achieved $(H(Z_1)/L , H(Z_2)/L , H(Z_3)/L) = (2, 2.5, 2)$. Note that the above scheme only specifies a judicious choice of the MDS matrices while all other assignment remains the same, so that the correctness and security of the scheme are not influenced.
While the individual randomness rate has been reduced for the above setting, it is not known if further saving is possible, i.e., optimality is open.
In general, understanding the fundamental limits of randomness cost and the potential tradeoff between communication cost and randomness cost for arbitrary $K, U, T$ is an interesting research direction for future study.

\subsection*{Non-identically Distributed Inputs}
We assume that the input vectors are uniform thus homogenous in the system model, while in federated learning, it is common that each user's data will be heterogeneous so that the input distributions will not be identical. Thus the secure sum computation problem with non-identically distributed inputs deserves further study. Note that our achievable scheme applies to arbitrarily distributed inputs (see Remark \ref{remark:noind}) while the converse proof critically relies on the uniformity of inputs. So uniform inputs are the worst case distribution (and a natural assumption if the distribution knowledge is not known or hard to acquire) and if we view uniform distributions as the baseline, the question is naturally how to incorporate the non-identical input distribution knowledge to produce a better (even optimal) scheme. Intriguingly, even if there is no user dropouts and we consider one-round communication protocols (this formulation reduces to one that has been extensively studied, see e.g., \cite{FKN, ishai1997private}), the secure sum computation problem depends on the structure of the input distribution in a subtle manner. For example, consider a two user problem, where $W_1 \in \textcolor{black}{\{5, 8, 10\}}, W_2 \in \{0, 2, 5\}$ and the server wishes to securely compute $W_1 + W_2$ over $\mathbb{F}_{11}$. For simplicity, the input length is assumed to be $L=1$. It turns out that the optimal communication strategy is for each user to send one symbol from $\mathbb{F}_7$, because the function $(W_1 + W_2)_{\mathbb{F}_{11}}$ has an invertible mapping as follows. 
{\begin{eqnarray}
\begin{array}{c|cccc}
(W_1 + W_2)_{\mathbb{F}_{11}} & 0 & 2 & 5 \\ \hline 
5 & 5 & 7 & 10 \\
8 & 8 & 10 & 2 \\
10 & 10 & 1 & 4 \\
\end{array} ~~\overset{\mbox{\scriptsize Invertible}}{\Longleftrightarrow}~~
\begin{array}{c|cccc}
(\tilde{W}_1 + \tilde{W}_2)_{\mathbb{F}_7} & 0 & 1 & 3 \\ \hline 
0 & 0 & 1 & 3 \\
2 & 2 & 3 & 5 \\
3 & 3 & 4 & 6 \\
\end{array}.
\end{eqnarray}
Therefore, we may set the messages as $X_1 = (\tilde{W}_1 + S)_{\mathbb{F}_7}, X_2 = (\tilde{W}_2 - S)_{\mathbb{F}_7}$, where $S$ is a noise variable that is uniform over $\mathbb{F}_7$ and independent of the inputs. In transforming the inputs \cite{Zhao_Sun_FKN}, the specific input structure is used in a non-trivial manner and interestingly, for this case, the scheme above is optimal for any joint distribution of the inputs as long as the probability is not zero over the support of $(W_1, W_2)$ because it matches a converse result from Theorem 3 of \cite{Data_Prabhakaran_Prabhakaran}. However, for arbitrary input distributions, the optimal communication cost for secure sum computation, to the best of our knowledge, is not known (even for two users). 

\subsection*{$\mathbb{F}_q$ versus $\mathbb{Z}_n$}
The input elements are assumed to be from the finite field $\mathbb{F}_q$ in this work, while inputs from the ring of integers modulo $n$, $\mathbb{Z}_n$ is also of interest. Note that when $n$ is not a prime number, $\mathbb{Z}_n$ defines a different algebraic object from $\mathbb{F}_q$. In federated learning, the input elements are typically integers quantized from real numbers and the sum operation is performed over $\mathbb{Z}_n$. While the converse bounds of this work generalize immediately to $\mathbb{Z}_n$ (see Remark \ref{remark:zn_con}), the achievable scheme relies on finite field operations, in particular, the generic property of Cauchy matrices. To cope with $\mathbb{Z}_n$, we may slightly modify the current scheme (with vanishing rate loss) as follows. The first round messages (see (\ref{eq:anss})) will be generated over $\mathbb{Z}_n$ (including the randomness variable $S_k$ and the sum operation), while the second round messages (see (\ref{eq:anss})) will be generated over $\mathbb{F}_q$ (including the randomness variables $N^{\mathcal{U}_1}$ and the matrix multiplication operation). This is possible because the second round messages are essentially (threshold ramp) secret shares of the sum of first round randomness $\sum_{k\in\mathcal{U}_1} S_k$ (referred to as the secret) and we can map the secret, which is an integer from $\mathbb{Z}_n$ due to the set-up of first round messages, to elements of the finite field $\mathbb{F}_q$ as long as $q > n$ (i.e., we can pick $q$ as the smallest prime power that is greater than $n$). The rate loss comes from the fact that $q > n$; however, through block codes, this penalty can be made arbitrarily small over long blocks, in terms of the message rate. Along similar lines, we can also stick to $\mathbb{Z}_n$ over the second round, where now we have to guarantee that the generic property of the precoding matrices (i.e., the Cauchy matrices in (\ref{eq:mds3}), which need to be replaced by other choices) as required by the achievability proof, is preserved. Essentially, we require various sub-matrices of the precoding matrices to have full rank and this issue has been studied in threshold ramp secret sharing literature, specifically over $\mathbb{Z}_{2^m}$ (e.g., quantize each input element to $2^m$ bits and sum under modulo $2^m$) \cite{Yamamoto, Kurihara_Kiyomoto_Fukushima_Tanaka}. Generic matrices (such as MDS matrices) over the integers have also been studied in recent work \cite{karingula2020codes}.

\subsection*{Uniform versus Pseudorandom Noise Variables}
We have focused exclusively on information theoretic security in this work, guaranteed by randomness variables that are functions of uniformly random noise variables. If we replace the uniform noise variables (also called random seeds) by pseudorandom variables that can be much shorter and are computationally indistinguishable from uniform random variables, then following standard cryptographic techniques \cite{Crypto_book, Crypto_Stinson}, the achievable scheme of this work can be shown to provide computational security with corresponding performance guarantees, albeit only for honest-but-curious adversaries (typically viewed as the weakest type of computational security guarantee in cryptography). Strengthening the power of the adversaries (e.g., who may actively deviate from the defined protocols) in our information theoretic security framework and connecting the solution to computational security is another interesting research direction for future study.

\subsection{Proof of Theorem \ref{theorem:random}: Total Randomness Converse}\label{sec:total}
We show that $H\left(\left( Z_k \right)_{k\in[K]} \right) \geq KL$. The proof is divided into four steps. 

Step 1: We show that $I\left(\sum_{k\in[K]} W_k; \left( X_k \right)_{k \in [K]} \right) = 0$.
Consider the security constraint (\ref{sec}) when $\mathcal{U}_1 = [K-1]$ and $\mathcal{T} = \emptyset$.
\begin{eqnarray}
0 &\overset{(\ref{sec})}{=}& I\left(\left(W_k\right)_{k\in[K]}; \left(X_k\right)_{k\in[K]}, \left(Y_k^{[K-1]}\right)_{k\in[K-1]} \Bigg| \sum_{k\in[K-1]} W_k \right) \\
&\geq& I\left( \sum_{k\in[K]} W_k ; \left(X_k\right)_{k\in[K]} \Bigg| \sum_{k\in[K-1]} W_k \right) \\
&=& I\left( \sum_{k\in[K]} W_k ; \left(X_k\right)_{k\in[K]}, \sum_{k\in[K-1]} W_k \right) - \underbrace{I\left( \sum_{k\in[K]} W_k ; \sum_{k\in[K-1]} W_k \right)}_{= 0} \\
&\geq&  I\left( \sum_{k\in[K]} W_k ; \left(X_k\right)_{k\in[K]} \right). \label{eq:s1}
\end{eqnarray}

Step 2: We show that $I\left( \left( W_k \right)_{k \in [K]} ; \left( X_k \right)_{k \in [K]} \right) = 0$. Consider the security constraint (\ref{sec}) when $\mathcal{U}_1 = [K]$ and $\mathcal{T} = \emptyset$.
\begin{eqnarray}
0 &\overset{(\ref{sec})}{=}& I\left(\left(W_k\right)_{k\in[K]}; \left(X_k\right)_{k\in[K]}, \left(Y_k^{[K]}\right)_{k\in[K]} \Bigg| \sum_{k\in[K]} W_k \right) \\
&\geq& I\left(\left(W_k\right)_{k\in[K]}; \left(X_k\right)_{k\in[K]} \Bigg| \sum_{k\in[K]} W_k \right) \\
&=&  I\left(\left(W_k\right)_{k\in[K]}, \sum_{k\in[K]} W_k; \left(X_k\right)_{k\in[K]} \right) -  \underbrace{ I\left( \sum_{k\in[K]} W_k ; \left(X_k\right)_{k\in[K]} \right) }_{\overset{(\ref{eq:s1})}{=} 0} \\
&\geq&  I\left(\left(W_k\right)_{k\in[K]}; \left(X_k\right)_{k\in[K]} \right). \label{eq:ss}
\end{eqnarray}

Step 3: We show that $H\left( \left(X_k\right)_{k\in[K]} \right) \geq KL$. To this end, we will use the message rate converse from the optimal rate characterization. Specifically, consider (\ref{eq:cr1}) and we have
\begin{eqnarray}
L &\overset{(\ref{eq:cr1})}{\leq}& H\left(X_{u} \big|  \left( W_k, Z_k \right)_{k\in[K]\backslash\{u\}} \right) \\
&\overset{(\ref{1x})}{=}& H\left(X_{u} \big|  \left( X_k, W_k, Z_k \right)_{k\in[K]\backslash\{u\}} \right) \\
&\leq& H\left(X_{u} \big|  \left( X_k \right)_{k\in[K]\backslash\{u\}} \right) \label{eq:s2}\\
\Rightarrow ~ H\left( \left(X_k\right)_{k\in[K]} \right) &=& \sum_{u=1}^K H\left(X_u \big| \left( X_k \right)_{k\in[u-1]} \right) \\
&\geq& \sum_{u=1}^K H\left(X_{u} \big|  \left( X_k \right)_{k\in[K]\backslash\{u\}} \right) \\
&\overset{(\ref{eq:s2})}{\geq}& KL. \label{eq:s3}
\end{eqnarray}

Step 4: We are now ready to show that $H\left(\left( Z_k \right)_{k\in[K]} \right) \geq KL$. 
\begin{eqnarray}
H\left(\left( Z_k \right)_{k\in[K]} \right) &\geq& H\left( \left( Z_k \right)_{k\in[K]} \big| \left( W_k \right)_{k\in[K]} \right) \\
&\overset{(\ref{1x})}{=}& H\left( \left( Z_k, X_k \right)_{k\in[K]} \big| \left( W_k \right)_{k\in[K]} \right) \\
&\geq& H\left( \left(X_k \right)_{k\in[K]} \big| \left( W_k \right)_{k\in[K]} \right) \\
&\overset{(\ref{eq:ss})}{=}& H\left( \left(X_k \right)_{k\in[K]} \right) \\
&\overset{(\ref{eq:s3})}{\geq}& KL.
\end{eqnarray}

\subsection{Proof of Theorem \ref{theorem:random}: Individual Randomness Converse}\label{sec:ind}
We show that when $U=1$, $H(Z_k) \geq KL, \forall k \in [K]$. To proceed, let us prove $H(Z_1) \geq KL$ and the proof of $H(Z_k) \geq KL, k \in [K]\backslash\{1\}$ follows from symmetry. The proof is divided into three steps.

Step 1: We show that $\forall k \in [K] \backslash \{1\}$, $I\left( Y_1^{\{1,k\}} ; W_k \big| \left(W_u \right)_{u\in[K]\backslash\{k\}}, \left(X_u\right)_{u\in[k-1]} \right) =0$. 
\begin{eqnarray}
L &=& \underbrace{I\left(W_1+W_k; \left(X_u\right)_{u\in[K]}, \left(Y_1^{\{1,k\}}\right) \right)}_{\overset{(\ref{corr})}{=} L} + \underbrace{I\left(\left(W_u\right)_{u\in[K]}; \left(X_u\right)_{u\in[K]}, Y_1^{\{1,k\}} \Big| W_1 + W_k \right)}_{\overset{(\ref{sec})}{=} 0} \label{eq:es3}
\\
&=& I\left(\left(W_u\right)_{u\in[K]}; \left(X_u\right)_{u\in[K]}, Y_1^{\{1,k\}} \right) \\
&=& I\left( Y_1^{\{1,k\}}, \left(X_u\right)_{u\in[k-1]}  ; \left(W_u \right)_{u\in[K]} \right) + I\left(  \left(X_u\right)_{u\in[K]\backslash [k-1]} ; \left(W_u \right)_{u\in[K]}  \Big| Y_1^{\{1,k\}}, \left(X_u\right)_{u\in[k-1]} \right) \notag\\
&&\\
&\overset{(\ref{corr})}{\geq}& I\left( Y_1^{\{1,k\}} ; \left(W_u \right)_{u\in[K]} \Big| \left(X_u\right)_{u\in[k-1]} \right)\notag\\
&& ~+ I\left(  \left(X_u\right)_{u\in[K]\backslash [k-1]}, W_1 + W_k ; \left(W_u \right)_{u\in[K]}  \Big| Y_1^{\{1,k\}}, \left(X_u\right)_{u\in[k-1]} \right)
\label{eq:es4}\\
&\geq& I\left( Y_1^{\{1,k\}} ; \left(W_u \right)_{u\in[K]} \Big| \left(X_u\right)_{u\in[k-1]} \right) + I\left( W_1 + W_k ; \left(W_u \right)_{u\in[K]}  \Big| Y_1^{\{1,k\}}, \left(X_u\right)_{u\in[k-1]} \right) \\
&=& I\left( Y_1^{\{1,k\}} ; \left(W_u \right)_{u\in[K]} \Big| \left(X_u\right)_{u\in[k-1]} \right) + H\left( W_1 + W_k  \Big| Y_1^{\{1,k\}}, \left(X_u\right)_{u\in[k-1]} \right) \\
&\geq& I\left( Y_1^{\{1,k\}} ; \left(W_u \right)_{u\in[K]} \Big| \left(X_u\right)_{u\in[k-1]} \right) + H\left( W_1 + W_k  \Big| Y_1^{\{1,k\}}, \left(W_u, Z_u, X_u\right)_{u\in[k-1]} \right) \\
&\overset{(\ref{1x}) (\ref{2y})}{=}& I\left( Y_1^{\{1,k\}} ; \left(W_u \right)_{u\in[K]} \Big| \left(X_u\right)_{u\in[k-1]} \right) + H\left( W_k  \Big| \left(W_u, Z_u\right)_{u\in[k-1]} \right) \label{eq:es5} \\ 
&\overset{(\ref{h1}) (\ref{h2})}{\geq}& I\left( Y_1^{\{1,k\}} ; W_k \Big| \left(W_u \right)_{u\in[K]\backslash\{k\}}, \left(X_u\right)_{u\in[k-1]} \right) + L \label{eq:ss1}
\end{eqnarray} 
where in (\ref{eq:es3}), the first term is $L$ because of the correctness constraint (\ref{corr}) when $\mathcal{U}_1 = [K], \mathcal{U}_2 = \{1\}$, i.e., $W_1 + W_k$ can be decoded from $(X_u)_{u\in[K]}, Y_1^{\{1,k\}}$ (the same argument is also used to obtain the second term of (\ref{eq:es4})) and the second term is zero because of the security constraint (\ref{sec}) when $\mathcal{U}_1 = [K], \mathcal{U}_2 = \{1\}$, i.e., the server can learn only $W_1 + W_k$. In (\ref{eq:es5}), the second term is due to the fact that messages $X_u, Y_u^{\mathcal{U}_1}$ are functions of $W_u, Z_u$. In (\ref{eq:ss1}), the second term follows from the independence of $(W_u)_{u\in[K]}$ and $(Z_u)_{u\in[K]}$, and the uniformity of $(W_u)_{u\in[K]}$.
After canceling $L$ on both hand sides of (\ref{eq:ss1}), we obtain the desired equation.

Step 2: We show that $\forall k \in [K] \backslash \{1\}$, $I\left( Z_1; X_k \big| \left(W_u \right)_{u\in[K]}, \left(X_u\right)_{u\in[k-1]} \right) \geq L$.
\begin{eqnarray}
&& I\left( Z_1; X_k \big| \left(W_u \right)_{u\in[K]}, \left(X_u\right)_{u\in[k-1]} \right) \notag\\
&\overset{(\ref{2y})}{=}& I\left( Y_1^{\{1,k\}}, Z_1; X_k \big| \left(W_u \right)_{u\in[K]}, \left(X_u\right)_{u\in[k-1]} \right) \label{eq:es1}\\
&\geq& I\left( Y_1^{\{1,k\}}; X_k \big| \left(W_u \right)_{u\in[K]}, \left(X_u\right)_{u\in[k-1]} \right) \\
&\overset{(\ref{eq:ss1})}{=}& I\left( Y_1^{\{1,k\}}; X_k, W_k \big| \left(W_u \right)_{u\in[K]\backslash\{k\}}, \left(X_u\right)_{u\in[k-1]} \right) \\
&\geq& I\left( Y_1^{\{1,k\}}; W_k \big| \left(W_u \right)_{u\in[K]\backslash\{k\}}, \left(X_u\right)_{u\in[k]} \right) \\
&=& H\left(W_k \big| \left(W_u \right)_{u\in[K]\backslash\{k\}}, \left(X_u\right)_{u\in[k]} \right) \notag\\
&&~- H\left( W_k \big| \left(W_u \right)_{u\in[K]\backslash\{k\}}, \left(X_u\right)_{u\in[k]}, Y_1^{\{1,k\}} \right) \\
&\overset{(\ref{eq:ss})}{=}& L - H\left( W_1 +W_k \big| \left(W_u \right)_{u\in[K]\backslash\{k\}}, \left(X_u\right)_{u\in[k]}, Y_1^{\{1,k\}} \right) \label{eq:es2}\\
&\overset{(\ref{corr})}{=}& L - 0 = L
\label{eq:ss2} 
\end{eqnarray}
where (\ref{eq:es1}) follows from the fact that $Y_1^{\{1,k\}}$ is a function of $W_1, Z_1$ (see (\ref{2y})), (\ref{eq:es2}) follows from the independence of $(X_k)_{k\in[K]}$ and $(W_k)_{k\in[K]}$ (see (\ref{eq:ss})) and the uniformity of $(W_k)_{k\in[K]}$ (see (\ref{h1}), (\ref{h2})), and to obtain (\ref{eq:ss2}) we use the correctness constraint (\ref{corr}) when $\mathcal{U}_1 = \{1,k\}, \mathcal{U}_2 = \{1\}$, i.e., $W_1 + W_k$ can be recovered with no error from $X_1, X_k, Y_1^{\{1,k\}}$.

Step 3: We are now ready to show that $H(Z_1) \geq K$.
\begin{eqnarray}
H(Z_1) &\geq& I\left( Z_1; \left( X_k \right)_{k\in[K]} \big| \left( W_k \right)_{k\in[K]} \right) \\
&=&  I\left( Z_1; X_1 \big| \left( W_k \right)_{k\in[K]} \right) + \sum_{k=2}^K I\left( Z_1; X_k \big| \left(W_u \right)_{u\in[K]}, \left(X_u\right)_{u\in[k-1]} \right)  \\
&\overset{(\ref{1x}) (\ref{eq:ss2})}{\geq}&  H\left( X_1 \big| \left( W_k \right)_{k\in[K]} \right) + (K-1)L \\
&\overset{(\ref{eq:ss})}{=}& H(X_1) + (K-1)L ~\overset{(\ref{eq:cr2})}{\geq}~ KL.
\end{eqnarray}

\section{Conclusion}
Motivated by secure aggregation in federated learning, we consider a secure sum computation problem with user dropouts and characterize the optimal communication efficiency under information theoretic security. This work represents a step towards using information and coding theory tools to understand diverse relevant challenges brought by new machine learning paradigms. 

\let\url\nolinkurl
\bibliographystyle{IEEEtran}
\bibliography{Thesis}
\end{document}